\documentclass[journal,twoside,web]{ieeecolor}
\usepackage{generic}
\usepackage{textcomp}

\usepackage[utf8]{inputenc}
\usepackage{hyperref}
\usepackage{float}
\usepackage{algorithm}
\usepackage[noend]{algpseudocode}
\usepackage{amssymb}
\usepackage{amsfonts}
\usepackage{nccmath}
\usepackage{mathtools}
\usepackage{float}
\usepackage[noadjust]{cite}
\usepackage{balance}
\usepackage[noabbrev]{cleveref}
\crefformat{equation}{(#1)}
\usepackage{bbm}
\usepackage{bm}
\usepackage{resizegather}
\usepackage[super]{nth}
\usepackage{acronym}
\usepackage{subcaption}
\usepackage{comment}

\usepackage{enumerate}
\let\labelindent\relax
\usepackage[inline]{enumitem}

\hypersetup{
    colorlinks=true,
    linkcolor=blue,
    filecolor=magenta,      
    urlcolor=cyan,
}

\setlist{  
  listparindent=\parindent,
  parsep=0pt,
}

\newtheorem{theorem}{Theorem}

\newtheorem{lemma}{Lemma}
\newtheorem{assumption}{Assumption}
\newtheorem{definition}{Definition}
\newtheorem{remark}{Remark}
\newtheorem{proposition}{Proposition}
\newtheorem{corollary}{Corollary}

\newcommand\x{\mbox{$\bm x$}}
\def\u{\mbox{$\bm u$}}

\newcommand \e{\mbox{$\bm e$}}

\newcommand\otheta{\mbox{$\bm \theta$}}
\newcommand\ophi{\mbox{$\bm \phi$}}
\newcommand\ozeta{\mbox{$\bm \zeta$}}

\newacro{DGR}{Data-Guided Regulation}
\newacro{F-DGR}{Fast Data-Guided Regulation}
\newacro{SVD}{Singular Value Decomposition}
\newacro{PBH}{Popov-Belevitch-Hautus}
\newacro{LQR}{Linear Quadratic Regulator}
\newacro{PI}{Policy Iteration}
\newacro{RLS}{Recursive Least-Squares}
\newacro{sysID}{system identification}
\newacro{LMI}{Linear Matrix Inequalities}
\newacro{PE}{Persistently Excited}
\newacro{ARE}{Algebraic Riccati Equation}
\newacro{MARL}{Multiagent Reinforcement Learning}
\newacro{MDP}{Markov Decision Processes}
\newacro{D2SPI}{Data-driven Structured Policy Iteration}
\newacro{SPE}{Subgraph Policy Evaluation}

\newcommand{\argmin}{\mathop{\rm arg\,min}}

\def\arrvline{\hfil\kern\arraycolsep\vline\kern-\arraycolsep\hfilneg}

\def\BibTeX{{\rm B\kern-.05em{\sc i\kern-.025em b}\kern-.08em
    T\kern-.1667em\lower.7ex\hbox{E}\kern-.125emX}}

\begin{document}
\title{Data-Driven Structured Policy Iteration for Homogeneous Distributed Systems}
\author{Siavash Alemzadeh$^\dagger$, Shahriar Talebi$^\dagger$, \IEEEmembership{Student Member, IEEE}, \\
and Mehran Mesbahi, \IEEEmembership{Fellow, IEEE}
\thanks{$^\dagger$S. Alemzadeh and S. Talebi contributed equally to this work.}
\thanks{The research of the authors has been supported by NSF grant SES-1541025, NSF grant ECCS-2149470, and AFOSR  grant  FA9550-20-1-0053.}
\thanks{This work was carried out while all authors were with the William E. Boeing department of Aeronautics and Astronautics at the University of Washington, Seattle. S. Alemzadeh is currently with Microsoft Corporation and S. Talebi is with the John A. Paulson School of Engineering and Applied Sciences at the Harvard University; emails: {\tt alems@uw.edu, shahriar@uw.edu, and mesbahi@uw.edu}.}
\thanks{A preliminary version of this work has been presented at the 60\textsuperscript{th} IEEE Conference on Decision and Control \cite{talebi2021distributed}. In comparison, the current manuscript expands on connections to broader related works, provides  proofs of all results, as well as including the suboptimality guarantees. Extended simulation results have also been included in this version of the manuscript.}
}

\maketitle

\begin{abstract}
  Control of networked systems, comprised of interacting agents, is often achieved through modeling the
underlying interactions.
Constructing accurate models of such interactions--in the meantime--can become prohibitive in applications.
{Data-driven control methods avoid such complications by directly synthesizing a controller from the observed data.}
{In this paper, we propose an algorithm} referred to as \acf{D2SPI}, for
synthesizing an efficient feedback mechanism {that respects the sparsity pattern induced by the underlying interaction network.}
{ In particular, our algorithm uses temporary ``auxiliary'' communication links in order to enable the required information exchange on a (smaller) sub-network during the ``learning phase''---links that will be removed subsequently for the final distributed feedback synthesis.}
We then proceed to show that the learned policy 
results in a stabilizing structured policy for the entire network.
Our analysis is then followed by showing the stability and convergence of the proposed distributed policies throughout the learning phase, exploiting a construct referred to as the ``Patterned monoid.''
The performance of \ac{D2SPI} is then
demonstrated using representative simulation scenarios.
\end{abstract}

\begin{IEEEkeywords}
Structured control, Patterned monoids, Data-driven Policy Iteration, Networked control systems
\end{IEEEkeywords}

\section{Introduction}
\label{sec:intro}
\IEEEPARstart{I}{n recent} years, there has been a renewed interest in distributed control of large-scale systems.
The unprecedented interdependence and
size of the data generated by such systems have
necessitated a distributed approach to policy
computation in order to influence or direct their behavior
and performance.
In these scenarios, collective actions are often synthesized 
via local decisions, informed by a {\em structured}
information exchange mechanism.
%
An important roadblock for centralized control design 
methods, is thereby,
their scalability and shortcomings in utilizing 
{the underlying structure of large-scale
interconnected systems.}\footnote{$\mathcal{O}(n^3)$ complexity of solving the \textit{Algebraic Riccati Equation} \cite{bini2011numerical} and scalability issues of \textit{Model Predictive Control} \cite{camponogara2002distributed} are among such examples.}

Structured control synthesis in the meantime is generally an NP-hard constrained optimization problem {\cite{papadimitriou1986intractable}}.
Hence, distributed control design for large-scale systems
has often been pursued
not necessarily to characterize 
optimal policies per se, but to devise
efficient (possibly suboptimal) control mechanisms that 
exploit the inherent system structure.
{In parallel, recent advances in measurement technologies have made available an unprecedented amount of data, motivating how offline and online data-processing can be leveraged for data-driven decision-making on high-dimensional complex systems.}

\begin{figure}[t]
    \centering
    \includegraphics[width=\columnwidth]{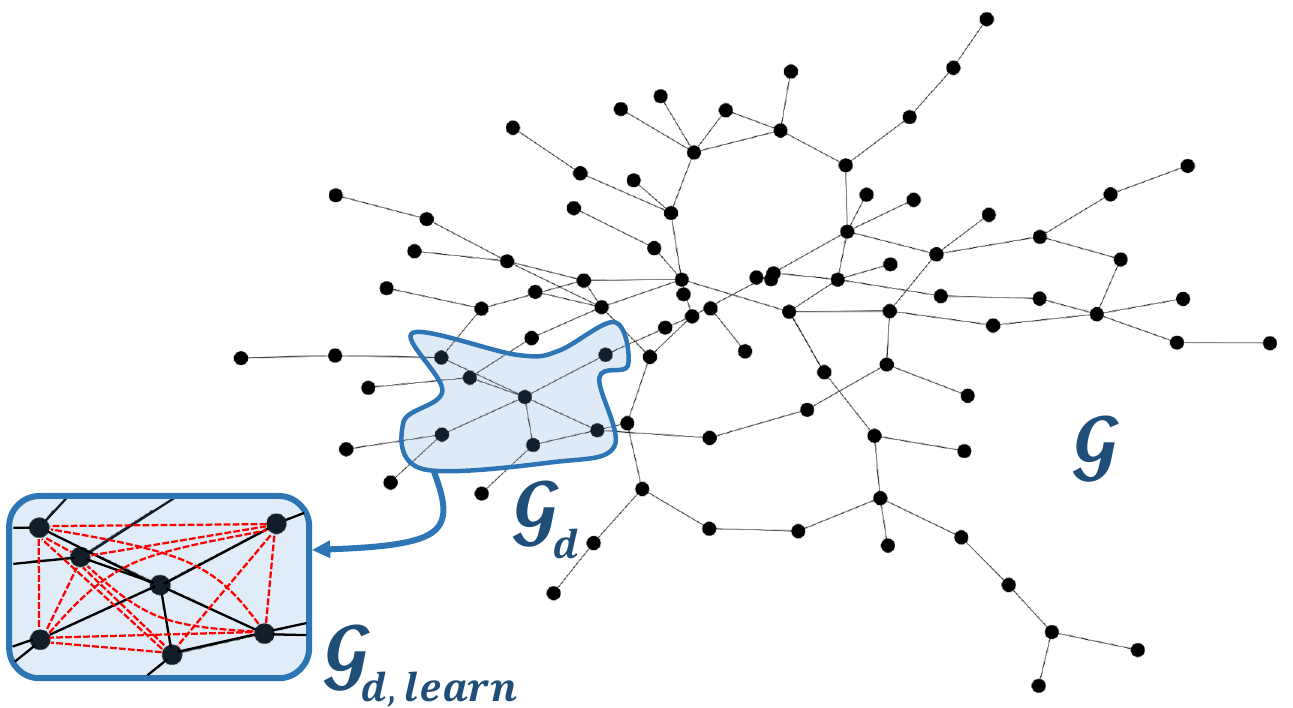}
    \caption{\small Addition of auxiliary links (dashed red) to the subgraph $\mathcal{G}_d$ during the policy learning phase. The size of the subgraph depends on the maximum degree of the original graph $\mathcal{G}$.}
    \label{fig:random_graph}
    \vspace{-0.5cm}
\end{figure}

In this work, we propose the linear-quadratic regulator (LQR)-based algorithm, coined \acf{D2SPI}, to iteratively learn stabilizing controllers for unknown but identical linear dynamical systems that are connected via
a network induced by the coupling in their performance.
The setup is a particular realization of cooperative 
game-theoretic decision-making (see remarks under \Cref{foot:game}).
This class of synthesis problems is motivated by applications such as formation flight \cite{stipanovic2004decentralized} and
distributed camera systems \cite{borrelli2005hybrid}, where 
the dynamics of the network nodes (agents)
cannot be {precisely} parameterized.
\ac{D2SPI} is built upon a data-driven learning
phase on a subgraph in a large network.
This subgraph includes the agent with
maximum degree in the network and requires {enabling}
\textit{auxiliary} links within this subgraph in order to iteratively learn a stabilizing structured controller {(optimal for the subgraph)} for the entire network--see \Cref{fig:random_graph}.
This ``extension'' synthesis procedure utilizes
a symmetry property of the networked systems,
that we refer to as \textit{Patterned monoid}
(see \Cref{subsec:motivation}).

The remainder of the paper is organized as follows.
In \S\ref{sec:problem_setup} we introduce the problem setup and motivation behind our work, and provide an overview of
the related literature (\S\ref{sec:relWork}).
In \S\ref{sec:main_algorithm}, we present and analyze the \ac{D2SPI} algorithm, followed by the theoretical analysis in \S\ref{sec:analysis}.
Illustrative examples are provided in \S\ref{sec:simulation}, followed by concluding remarks in \S\ref{sec:conclusion}. 

\textbf{Notation.} 
The operator $\mbox{diag}(\cdot)$ makes a square diagonal matrix out of the elements of its argument; $\mbox{vech}(\cdot)$ on the other hand, takes a square matrix and stacks the lower left triangular half (including the diagonal) into a single vector.
{We use $N\succ0$ ($\succeq0$) to declare $N$ as a positive-(semi)definite matrix.}
The $i$th eigenvalue and spectral radius of $M$ are denoted by $\lambda_i(M)$ and $\rho(M)$; $M$ is called Schur stable when
$\rho(M)<1$.
We say that an $n$-dimensional linear system parameterized 
by the pair $(A,B)$ is \emph{controllable} if the controllability matrix $\mathcal{C}=[B\quad AB\ \dots\ A^{n-1}B]$ has a full-rank.
We denote the \emph{Kronecker product} of two matrices by $\otimes$.
{For a block matrix $\mathbf{F}$, by $[\mathbf{F}]_{r k}$ we imply the $r$th row and $k$th column ``block'' component with appropriate dimensions.}
An (undirected) graph is characterized by $\mathcal{G}=(\mathcal{V}_\mathcal{G},\mathcal{E}_\mathcal{G})$ where $\mathcal{V}_\mathcal{G}$ is the set of nodes and $\mathcal{E}_\mathcal{G}\subseteq \mathcal{V}_\mathcal{G}\times\mathcal{V}_\mathcal{G}$ denotes the set of edges.
An edge exists from node $i$ to $j$ if (the unordered pair) $(i,j)\in \mathcal{E}_\mathcal{G}$; this is also specified by writing $j\in\mathcal{N}_i$, where $\mathcal{N}_i$ is the set of neighbors of node $i$.
{We designate the maximum degree of $\mathcal{G}$ by $d_{\max}(\mathcal{G})$.}
{Finally, the graph $\mathcal{G}$ can be represented 
using matrices such as the Laplacian $\mathcal{L}_\mathcal{G}$ or the {adjacency} $\mathcal{A}_\mathcal{G}$. To distinguish between system dynamics quantities related to the entire graph and a subgraph, we utilize hat and tilde notation, respectively.}
{ A \textit{semigroup} is a set and a binary operator in which the multiplication operation is associative (but its elements need not have inverses). A \textit{monoid} is a semigroup with an identity element. A \textit{group} is a monoid each of whose elements is invertible.
We denote the set of symmetric $n\times n$ real matrices by $\mathbb{S}^{n}$, and the set invertible ones by $\mathrm{G L}(n,\mathbb{R})$---which is a group under matrix multiplication also known as the general linear group.
}
\section{Problem Setup}
\label{sec:problem_setup}

Consider a network of identical agents with interdependencies induced by
a network-level objective.
In particular, we assume that the system contains $N$ agents forming a graph $\mathcal{G}=(\mathcal{V}_\mathcal{G},\mathcal{E}_\mathcal{G})$, where each node of the graph in $\mathcal{V}_\mathcal{G}$ represents a linear discrete-time system, 
\begin{equation}
    \begin{aligned}
        \label{eq:agent_i_dynamics}
    	\x_{i,t+1} = A \x_{i,t} + B \u_{i,t},
    	\hspace{10mm} i = 1,2,\dots,N,
    \end{aligned}
\end{equation}
with $\x_{i,t}\in\mathbb{R}^{n}$ and $\u_{i,t}\in\mathbb{R}^{m}$ denoting the state and input of agent $i$ at time-step $t$, respectively. 
The unknown system matrices $A\in\mathbb{R}^{n\times n}$ and $B\in\mathbb{R}^{n\times m}$ 
are assumed to form a controllable pair.
The network dynamics can compactly be represented as,
\begin{align}
    \label{eq:dynamics_N_agents}
    \hat{\mathbf{x}}_{t+1} = \hat{\mathbf{A}} \hat{\mathbf{x}}_t + \hat{\mathbf{B}} \hat{\mathbf{u}}_t,
\end{align}
where $\hat{\mathbf{x}}_t\in\mathbb{R}^{Nn}$ and $\hat{\mathbf{u}}_t\in\mathbb{R}^{Nm}$ are comprised
of the states and inputs of entire network,
\(
    \hat{\mathbf{x}}_t = \big[ \x_{1,t}^\intercal \, \dots \, \x_{N,t}^\intercal \big]^\intercal, \quad \hat{\mathbf{u}}_t = \big[ \u_{1,t}^\intercal \, \dots \, \u_{N,t}^\intercal \big]^\intercal,
\)
with $\hat{\mathbf{A}}\in\mathbb{R}^{Nn\times Nn}$ and $\hat{\mathbf{B}}\in\mathbb{R}^{Nn\times Nm}$ are in block diagonal forms $\hat{\mathbf{A}} = \mathrm{I}_{N} \otimes A$ and $\hat{\mathbf{B}} = \mathrm{I}_{N} \otimes B$.
The agents' interconnections are represented
by  edges  $\mathcal{E}_\mathcal{G}\subseteq \mathcal{V}_\mathcal{G}\times\mathcal{V}_\mathcal{G}$ that 
can facilitate a distributed feedback design.
We do not assume that $\mathcal{G}$ is necessarily connected; the motivation for this becomes apparent subsequently.
Let $\mathcal{N}_i$ denote the set of neighbors of node $i$ in $\mathcal{G}$ (excluding itself).
Then, based on the underlying communication graph and for any choice of positive integers $a$ and $b$, we define a linear subspace of { real-valued ${a N \times b N}$ matrices} as,
\begin{gather*}
    \begin{aligned}
    \quad \mathcal{U}_{a,b}^N(\mathcal{G}) \coloneqq \big\{ \mathbf{M} \in \mathbb{R}^{a N \times b N}& \vert\ [\mathbf{M}]_{ij} = \mathbf{0}\ \text{if}\ j\not\in\mathcal{N}_i\cup\{i\}, \\ &[\mathbf{M}]_{ij} \in \mathbb{R}^{a\times b},\ i,j=1,\cdots,N \big\}.
\end{aligned}
\end{gather*}

Without having access to the system parameters $A$ and $B$, we are interested in designing linear feedback gains, consistent with the desired sparsity pattern induced by the network, using data generated by~\eqref{eq:dynamics_N_agents}.
More precisely, given an initial condition $\hat{\mathbf{x}}_1$, the distributed (structured) optimal control problem assumes the form,
\begin{equation}
    \begin{aligned}
        \label{eq:distributed_optimization}
        &\textstyle \hspace{3mm}\min_{\hat{\mathbf{K}}} \hspace{3mm} \sum_{t=1}^{\infty}  \hat{\mathbf{x}}_t^{\intercal} \hat{\mathbf{Q}} \hat{\mathbf{x}}_t + \hat{\mathbf{u}}_t^{\intercal} \hat{\mathbf{R}} \hat{\mathbf{u}}_t \\
        &\hspace{5mm}\text{s.t.}\hspace{4mm} \eqref{eq:dynamics_N_agents}, \quad \hat{\mathbf{u}}_t = \hat{\mathbf{K}} \hat{\mathbf{x}}_t, \quad \hat{\mathbf{K}} \in \mathcal{U}_{m,n}^N(\mathcal{G}),
    \end{aligned}
\end{equation}
where $\hat{\mathbf{K}}$ stabilizes the pair $(\hat{\mathbf{A}}, \hat{\mathbf{B}})$ (i.e., $\rho(\hat{\mathbf{A}} + \hat{\mathbf{B}}\hat{\mathbf{K}}) < 1$), $\hat{\mathbf{R}} = \mathrm{I}_N \otimes R$ and $\hat{\mathbf{Q}} = I_N \otimes Q_1 + \mathcal{L}_\mathcal{G} \otimes Q_2$ for some given cost matrices $Q_1\succ 0$, $Q_2\succeq 0$, $R\succ 0$.
Note that $\hat{\mathbf{Q}} \in \mathcal{U}_{n,n}^N(\mathcal{G})$ is positive definite.
Such interdependence induced through the
cost has been considered in a graph-based distributed control framework; see for instance \cite{borrelli2008distributed,deshpande2011distributed,wang2017distributed,massioni2009distributed}.
In a nutshell, the first term in $\hat{\mathbf{Q}}$ encodes the cost pertinent to state regulation for each agent, while the second term, captures the 
``disagreement'' cost between the neighbors.\footnote{\label{foot:game}One instance of such an interactive cost among agents appears in the cooperative game setup where agent $i$ aims to solve the minimization problem,
\begin{multline*}
    \min_{(\u_{i,t})_{t=0}^\infty \in \ell_2}\mathbf{J}_i(\hat{\mathbf{x}}_1, \hat{\mathbf{u}}_t) = N\sum_{t=0}^{\infty} \bigg( \x_{i,t}^\intercal Q_1 \x_{i,t} + \u_{i,t}^\intercal R \u_{i,t} \\
    + \sum_{j\in\mathcal{N}_i} \big( \x_{j,t} - \x_{i,t} \big)^{\intercal} Q_2 \big( \x_{j,t} - \x_{i,t} \big) \bigg).
\end{multline*}
Then, it is well-known that the set of Pareto front solution of this game can be obtained by minimizing the parametric cost function,
\begin{align*}
    \min_{(\u_{1,t})_{t=0}^\infty, \dots, (\u_{N,t})_{t=0}^\infty \in \ell_2}\ \sum_{i=1}^N \alpha_i \mathbf{J}_i(\hat{\mathbf{x}}_1, \hat{\mathbf{u}}_t),
\end{align*}
parameterized by $\alpha_1,\dots,\alpha_N$ where $\alpha_i \in [0,1]$ and $\sum_i \alpha_i = 1$ (see e.g. \cite{engwerda2005lq}).
Therefore, a cost such as in \eqref{eq:distributed_optimization} can be viewed as a special case of the fair Pareto optimal solution with the choice of $\alpha_i= 1/N$ for all $i$.}

In this work, we propose 
a data-guided suboptimal solution for \eqref{eq:distributed_optimization}, not
relying on knowledge of the system parameters $\hat{\mathbf{A}}$ and $\hat{\mathbf{B}}$.
Instead, our approach relies on the system's input-state time series for
synthesizing distributed feedback control 
on $\mathcal{G}$.
A summary of challenges in analyzing this problem is listed as follows:
\begin{enumerate}
    \item
    {
    The constrained optimization problem in \eqref{eq:distributed_optimization} is in general NP-hard~\cite{papadimitriou1986intractable, blondel1997np}. 
    In particular, the problem of stabilization by decentralized static state feedback is NP-hard if one imposes a bound on the norm of the controller \cite{blondel1997np}. Even though the result is not shown for the case with no bound on the controller parameters, the general problem is commonly believed to be a computationally hard problem. Note that stabilization is a necessity for feasibility of optimization problem in (3) for arbitrary initial state $\mathbf{x}_0$ whenever $Q_1 \succ 0$. This is simply because the cost is unbounded for an unstable policy.
    Nonetheless, based on the complete knowledge of the system parameters, this problem has been investigated under variety of assumptions \cite{gupta2005sub, rotkowitz2005characterization, bamieh2002distributed, borrelli2008distributed}, or approached directly with the aid of projected gradient-based policy updates \cite{maartensson2009gradient, bu2019lqr}.
    }
    
    \item  In the meantime, policies obtained via data-driven approaches, do not necessarily respect the hard constraints on $\hat{\mathbf{K}} \in \mathcal{U}_{m,n}^N(\mathcal{G})$ posed in \eqref{eq:distributed_optimization}.
    In particular, we point out that ``projection'' onto the intersection of the constraint imposed by the network and stabilizing controllers is not straightforward due to the intricate geometry of the set of stabilizing controllers \cite{bu2020topological}.
    
    \item Another key challenge in adopting data-driven methods for the entire network 
    is rooted in the ``curse of dimensionality'' inherent in the design and analysis of large-scale systems. In fact, even collecting data from the entire network can be prohibitive.
    
    \item  Finally, it is often impossible in  applications to pause the operation of the network for data collection or decision-making purposes.\footnote{For example, consider an operational large-scale homogeneous aerial vehicles that need to coordinate their relative states (in addition to their respective state regulation) over a network induced by their proximity~\cite{stipanovic2004decentralized}.}
\end{enumerate}

\subsection{Structures in the Problem and our Approach}
\label{subsec:motivation}
{

In this work the sparsity requirement $\hat{\mathbf{K}} \in \mathcal{U}_{m,n}^N(\mathcal{G})$ is considered as a \emph{hard constraint} for control synthesis, and as such, the corresponding optimization is challenging in general. Hence, we shift our attention from the optimal solution of \eqref{eq:distributed_optimization} towards a ``reasonable'' suboptimal stabilizing distributed controller with a scalable computational cost. 
Henceforth, we aim to exploit the problem structure that is incurred due to the homogeneity of system dynamics and the pattern in performance index respecting the underlying graph topology---see the definition of $\hat{\mathbf{Q}}$ depending on the graph Laplacian $\mathcal{L}_\mathcal{G}$ in \cref{eq:distributed_optimization}. Additionally, as the system dynamics parameters in \cref{eq:dynamics_N_agents} are unknown, we adapt a $\mathcal{Q}$-learning procedure to our setup that provably converges to a distributed policy with suboptimality guarantees.

In the \emph{absence} of the sparsity constraint  $\hat{\mathbf{K}} \in \mathcal{U}_{m,n}^N(\mathcal{G})$, a learning approach for solving the optimal control problem  \cref{eq:distributed_optimization} is the well-known $\mathcal{Q}$-learning procedure that was first introduced by Bradtke in \cite{bradtke1994adaptive,bradtke1994incremental}. Adopting this approach for (\ref{eq:distributed_optimization}) would require utilizing the input-state data trajectories of the entire networked systems in \cref{eq:dynamics_N_agents} to implement a quasi-Newton method for iteratively updating $\hat{\mathbf{K}}$ introduced by Hewer in \cite{hewer1971iterative}. If the system is controllable, Hewer's algorithm converges to the globally optimal solution with a quadratic rate, and so does the Bradtke's algorithm if in addition the data trajectories are ``informative'' enough---this is usually satisfied by a sufficient condition on the input signal referred to as ``persistence of excitation.''  
The main issue with this approach is the fact that the policy obtained in this way, will not respect the hard constraint of $\hat{\mathbf{K}} \in \mathcal{U}_{m,n}^N(\mathcal{G})$ as was posed in the optimization \eqref{eq:distributed_optimization}. 
This issue is particularly critical when a projection of the iterated policy on the set $\mathcal{U}_{m,n}^N(\mathcal{G})$ is not practical, or even fails to be stabilizing.
Additionally, collecting data trajectory from entire networked system in \cref{eq:dynamics_N_agents} can be expensive.

Inspired by the $\mathcal{Q}$-learning approach, we propose a model-free structured policy iteration scheme for the synthesis problem~\eqref{eq:distributed_optimization} with iterative stability, convergence, and reasonable performance guarantees. 
While the detailed algorithm is presented subsequently in \Cref{sec:main_algorithm}, in what follows we summarize the key steps of our design procedure for distributed data-driven policy iteration:
\begin{enumerate}
    \item Inherent to our distributed learning algorithm is a synthesis sub-problem whose dimension is related only to the maximum degree in underlying graph rather than the dimension of the original network.
    In particular, we will reason that for the learning phase, our method only requires data collection from a
    (specific) smaller sub-network $\mathcal{G}_d \subseteq \mathcal{G}$ with size $d = d_{\max}(\mathcal{G})+1$.
    This subgraph is substantially smaller than the original graph whenever $d_{\max}(\mathcal{G})$ is significantly smaller than $N$, reflecting the empirical feature of many real-world networks. In the meantime, our approach requires adding temporary communication links to the subgraph $\mathcal{G}_d$ to make it a clique during the learning phase-- we will discuss why this learning phase clique is required in Step 3 below---see \Cref{fig:random_graph}. The additional links are subsequently removed for the final data-driven feedback design. In robotic applications, such a clique can be established by temporary increasing the transmission range for neighboring nodes of the agent with maximum degree. Since longer range information transfer is generally costly, power levels for these agents are ``dialed back'' to their original settings following the learning phase.
    \item Then, we adapt the $\mathcal{Q}$-learning technique in order to learn a specific optimal policy for this sub-problem using data only from the systems in the subgraph $\mathcal{G}_d$. In particular, the proposed algorithm learns an optimal policy for the subgraph with a (off-)diagonal pattern consisting of two distinct system-level policies $K^*$ and $L^*$, representing the ``individual'' and ``cooperative'' components, respectively.
    More specifically, for any integer $r \geq 2$, 
    we can define a linear subspace of $\mathbb{R}^{rn \times rn}$ as
    \begin{multline*}
         \mathrm{L}(r\times n, \mathbb{R}) \coloneqq \big\{\mathbf{N}_r \in \mathbb{R}^{rn\times rn} \, \big| \, \mathbf{N}_r =\\
         \mathrm{I}_r \otimes ( A - B ) + \mathbbm{1}_r \mathbbm{1}_r^\intercal \otimes B, \; \text{for some } A, B  \big\},
    \end{multline*}
    which is also closed under matrix multiplication.
    Then, the policy learned for the subgraph $\mathcal{G}_d$ takes the form
    $\widetilde{\mathbf{K}}^* = \mathrm{I}_d \otimes \big( K^* - L^* \big) + \mathbbm{1} \mathbbm{1}^{\intercal} \otimes L^*$ which also implies that the closed-loop system $\widetilde{\mathbf{A}}_{\widetilde{\mathbf{K}}^*} \coloneqq \widetilde{\mathbf{A}} + \widetilde{\mathbf{B}} \widetilde{\mathbf{K}}^*$ lies in $\mathrm{L}(d\times n, \mathbb{R})$. This is particularly useful from a design perspective as it becomes clear in the next step.
  
    \item Next, we define the \textit{Patterned monoid} as\footnote{The proposed Patterned monoid and characterizing its interplay with the Lyapunov equation is considered as a key contribution in our approach. Note that, with reference to \cite{talebi2021distributed}, $\mathrm{PM}(r\times n, \mathbb{R})$ requires further qualifications to be a linear subgroup of $GL(rn,\mathbb{R})$.The monoid characterization is sufficient for our purposes and further extensions are deferred to our future work.}
    \begin{multline*}
     \mathrm{PM}(r\times n, \mathbb{R}) \coloneqq  \big\{\mathbf{N}_r \in \mathrm{L}(r\times n, \mathbb{R}) \cap \mathrm{G L}(rn,\mathbb{R}) \; \big|\; \\ 
      \text{ for some } A \in \mathrm{G L}(n,\mathbb{R}) \cap \mathbb{S}^{n}, \; B \in \mathbb{S}^{n} \big\}.
    \end{multline*}
    We note that the Patterned monoid $\mathrm{PM}(r\times n, \mathbb{R})$ is indeed a sub-monoid of {$\mathrm{G L}(rn,\mathbb{R})$}---following by the proof of \Cref{lem:tools_for_analysis}---i.e., it is closed under matrix multiplications and contains the identity matrix.
    The next observation underscores the relevance of Patterned monoids in system analysis.
    \begin{proposition}\label{prop:lineargroup}
    For a Schur stable matrix $A \in \mathrm{L}(r\times n, \mathbb{R})$ and $0 \prec Q \in \mathbb{S}^{rn}$, let $P$ denote the unique solution to the corresponding discrete-time Lyapunov equation, i.e., $P = A^\intercal P A + Q$. Then, $P \in \mathrm{PM}(r\times n, \mathbb{R})$ if and only if $Q \in \mathrm{PM}(r\times n, \mathbb{R})$.
    \end{proposition}
    
    The invariance of the Lyapunov equation under
    the action of Patterned monoid has
    important implications for our data-driven synthesis
    for large-scale networks---see \Cref{prop:H_structure}.
Another key ingredient of our approach, motivated by utilizing the
Patterned monoid in the context of $\mathcal{Q}$-learning, involves 
 allowing temporary communication links on the subgraph $\mathcal{G}_d$ in order to make it a clique only \emph{during the learning phase}---links that are subsequently removed. It then follows that the state cost parameter $\widetilde{\mathbf{Q}}$ for the subgraph $\mathcal{G}_d$ lies in $\mathrm{PM}(d\times n, \mathbb{R})$. Then, as the obtained closed-loop system $\widetilde{\mathbf{A}}_{\widetilde{\mathbf{K}}}$ at each iteration of the learning phase lies in $\mathrm{L}(d\times n, \mathbb{R})$, \Cref{prop:lineargroup} would imply that the associated cost matrix $\widetilde{\mathbf{P}}_{\widetilde{\mathbf{K}}}$ must lie in $\mathrm{PM}(d\times n, \mathbb{R})$ with two components $P_1 \in \mathrm{G L}(n,\mathbb{R}) \cap \mathbb{S}^{n}$ and $P_2 \in \mathbb{S}^{n}$.

    \item Next, an iterative $\mathcal{Q}$-learning procedure is designed based on both the patterned structure of the policy with system-level components $K$ and $L$, and the cost matrix with $P_1$ and $P_2$ components. This procedure provably converges to an optimal policy for the subgraph $\mathcal{G}_d$--see \Cref{thm:convergence}.

    \item We note that terminating the operation of the entire networked system for the purpose of data collection/learning is often infeasible in real-world applications. For example, disrupting the operation of power generators and loads for improving their respective network-level performance is highly undesirable. 
    Therefore, in addition to learning the optimal policy for the subgraph, in this work, we aim to simultaneously devise and update a policy for the rest of the network. 
    In this direction, our algorithm iteratively learns a ``stability margin'' $\tau$ according to the learned components of the policy at each iteration--see \Cref{prop:gain_margin}. This feature, together with homogeneity of the network, facilitates extending the policy synthesis procedure to a stabilizing policy for the entire network by utilizing the individual and cooperative components learned from the subgraph $\mathcal{G}_d$--see \Cref{thm:stabilizability_learning_phase}. 

    \item After the $\mathcal{Q}$-learning procedure has converged, the learning phase is terminated. The framework now allows removing the temporary links added to $\mathcal{G}_d$ during the ``clique subgraph learning phase" and the stability of the final learned policy will be guaranteed--see \Cref{cor:ultimate_stabilizability}.

    \item Finally, note that the learned policy for the entire graph will be, in general, a suboptimal solution to the optimization in \Cref{eq:distributed_optimization}. Yet, we provide guarantees on its sub-optimality gap in \Cref{thm:suboptimality} and illustrate its numerical performance in \Cref{sec:simulation}.
    
\end{enumerate}

The distributed control underpinning of the method proposed in this work follows its model-based analogues studied in~\cite{borrelli2008distributed, wang2017distributed}. In this work, our aim is to extend these approaches and propose a model-free structured policy iteration algorithm, which is not only computationally efficient, but also practical for operational large-scaled networked systems.
}


\subsection{Related Literature}
\label{sec:relWork}

 Distributed control is a well-established area of research in systems theory.
The roots of the field trace back to the socioeconomics literature in 1970's \cite{mcfadden1969controllability} and early works in the control literature followed suite later during that decade~\cite{wang1973stabilization}.
The main motivation for these
works was lack of scalability 
in centralized planning and control, 
due to information or computational 
limitations~\cite{sandell1978survey, ioannou1986decentralized}.
Fast forward a few decades, sufficient graph-theoretic conditions were provided for stability of formations comprised of identical vehicles \cite{fax2004information} and, along the same lines, graph-based distributed controller synthesis was further examined independently in works such as \cite{massioni2009distributed, deshpande2012sub, borrelli2008distributed, maartensson2009gradient}.
The topic was also studied from 
the perspective of spatial invariance~\cite{bamieh2002distributed, motee2008optimal} and a compositional layered design \cite{chapman2017data, alemzadeh2018influence}.
Moreover, from an agent-level perspective, the problem has been tackled for both homogeneous systems \cite{massioni2009distributed, borrelli2008distributed, wang2017distributed} and more recently heterogeneous ones \cite{sturz2020distributed}.

Having access to the underlying system model is a common assumption in the literature on distributed control, where the goal is to find a distributed feedback mechanism that
conforms to an underlying network topology.
However, deriving dynamic models from first principles could be restrictive for large-scale systems and complex mission scenarios~\cite{vasisht2019data}.
Such restrictions also hold for parametric perturbations that occur due to inefficient modeling or other unknown design factors.
For instance, even the LQR solution with its strong input robustness properties, may have small stability margins for general parameter perturbations \cite{dorato1994linear}.
Robust synthesis approaches could alleviate this issue when the perturbations follow specific models, in both centralized \cite{khargonekar1990robust} and distributed \cite{li2012distributed} cases.
However, if the original estimates of system parameters are inaccurate or the perturbations violate the presumed model, then both stability and optimality of the proposed feedback mechanisms 
can be compromised.
Data-driven control, on the other hand, circumvents such drawbacks and utilize
the available data generated by the system
when its model is unavailable.
This point of view has historically been examined in the
context of adaptive control and system identification \cite{ljung1999system}, particularly, when asymptotic properties of the
synthesized system are of interest.
For more recent works that have adopted a 
non-asymptotic outlook on data-driven control,
we mention~\cite{van2020data, alaeddini2018linear, dean2019sample} that used batched data 
for synthesis, as well as 
online iterative procedures~\cite{talebi2020online, oymak2019non}.
Furthermore, in regards to the adaptive nature of such algorithms, there is a close connection between online data-driven control and reinforcement learning \cite{lewis2012reinforcement, bradtke1994adaptive}.
In these latter works, policy iteration has been extended to approximate LQR by avoiding the direct solution of \ac{ARE}; yet majority of these works do not have favorable scaling
properties.

Control and estimation for large-scale systems offers its unique set of challenges due higher levels of uncertainty, scalability issues, and modeling errors.
Nevertheless, model-free synthesis for large-scale systems, as a discipline, is still in its infancy.
From a control theoretic perspective, the work~\cite{luo2019natural} addresses some of the aforementioned issues using ideas from mean-field multiagent systems and with the key assumption of partial exchangeability.
The work~\cite{alemzadeh2019distributed} on the other hand, provides a decentralized LQR algorithm based on network consensus that has low complexity, but potentially a high cost of implementation.
Lastly, SDP projection-based analysis has
been examined in \cite{chang2021distributed}, where each agent will have a sublinear regret as compared 
with the best fixed controller in hindsight.
The problem has also been considered from a game-theoretic standpoint \cite{li2017off, talebi2019distributed, nowe2012game},
where agents can have conflicting objectives.

{ In the following, we propose our algorithm that iteratively learns necessary control components from a sub-network that would be used to design a distributed controller for the entire network. We also show that depending on the structure of the problem, not only would our scheme is computationally efficient, but also more applicable when model-based control in high dimensions is (practically) infeasible. The structure of our distributed control design is inspired by~\cite{borrelli2008distributed}, that was subsequently extended to discrete-time in \cite{wang2017distributed}.}


\section{Main Algorithm}
\label{sec:main_algorithm}

In this section, we present and discuss the main algorithm of the paper, namely, \ac{D2SPI}.
Given the underlying communication graph $\mathcal{G}$, the networked system is considered as a black-box,
whereas the designer is capable of injecting input signals to the system and observe the corresponding states.
The goal of \ac{D2SPI} is then to find a data-guided suboptimal solution for \eqref{eq:distributed_optimization} without knowledge of system parameters $\hat{\mathbf{A}}$ and $\hat{\mathbf{B}}$.
To this end, our approach involves 
considering the synthesis problem 
on a subgraph $\mathcal{G}_d \subseteq \mathcal{G}$,
with the associated time-series data.
Before presenting the main algorithm, we formalize two useful notions in order to facilitate the presentation.
\begin{definition}
    \label{def:policy}
    Given a subgraph $\mathcal{G}' \subseteq \mathcal{G}$ and a node labeling, let $\mathrm{Policy} \left( \mathcal{V}_{\mathcal{G}'} \right)$ denote the concatenation of policies of the agents in $\mathcal{V}_{\mathcal{G}'}$, i.e.,
    \(
        \mathrm{Policy} \left( \mathcal{V}_{\mathcal{G}'} \right) \coloneqq [ \u_1^\intercal~~\u_2^\intercal~~\cdots~~\u_{|\mathcal{V}_{\mathcal{G}'}|}^\intercal ]^\intercal,
    \)
    where $\u_i$ is the feedback control policy of agent $i$ in the subgraph $\mathcal{G}'$ as a mapping from $\{\x_j|j\in\mathcal{N}_i\cup\{i\}\}$ to $\mathbb{R}^m$. Furthermore, we use $\mathrm{Policy}( \mathcal{V}_{\mathcal{G}'}) |_t$ to denote the realization of these policies at time $t$. Similarly, we define,
    \(\mathrm{State} \left( \mathcal{V}_{\mathcal{G}'} \right) \coloneqq [\x_1^\intercal~~\x_2^\intercal~~ \cdots~~\x_{|\mathcal{V}_{\mathcal{G}'}|}^\intercal]^\intercal.\)
\end{definition}
The \ac{D2SPI} algorithm is introduced in \Cref{alg:distributed_control_algorithm} with the following standard assumption.
\begin{assumption}\label{assmp:main}
The initial controller $K_1$ is stabilizing for the controllable pair $(A,B)$, and $\e_t$ in \Cref{alg:SPE} is such that $\mathrm{Policy}(\mathcal{V}_{\mathcal{G}'}) |_t$ remains persistently exciting (PE); 
{
more precisely, if we collect the state and input difference signals in the vector form $\ophi_t$ as defined in \Cref{alg:SPE}, then the PE condition requires that \cite{bradtke1994incremental}: there exists a positive integer $N_0$ and positive constants $\epsilon_0 \leq \overline{\epsilon}_0$ such that
\[\epsilon_0 I \leq \frac{1}{N}\sum_{i=1}^N \ophi_{t-i} \ophi_{t-i}^\intercal \leq \overline{\epsilon}_0 I, \; \text{for all} \; t\geq N_0 \text{ and } N\geq N_0.\]
}
\end{assumption}

\begin{remark}
    
    Note that the controllability of the pair $(A,B)$ and the PE condition are common sufficient assumptions in data-driven control literature \cite{goodwin2014adaptive,Willems2005note}. 
    The PE condition above is an adaptation of ``strong persistent excitation'' in \cite{goodwin2014adaptive} to the least-squares problem of recovering unknown parameters of the so-called $\mathcal{Q}$-function. The PE condition can be easily satisfied in our setup by ensuring a rich randomness in the signal $\e_t$, such as a non-degenerate Guassian distribution.\\
    A more modern treatment of PE condition is stated by a rank condition on the input Hankel matrices in the context of Willems' Fundamental Lemma \cite{Willems2005note, yue2021controllability}.
    We also refer to \cite{yue2021controllability} for a discussion on how these assumptions can be relaxed. Furthermore, assuming initial stabilizing controller $K_1$ is also common in the policy iteration literature.  For instance, in the case of open loop stable system $A$, $K_1$ is simply chosen to be zero. For an unknown and unstable pair $(A,B)$, more elaborate online stabilization techniques have been adopted that are out of the scope of this work. For one such method we refer to \cite{talebi2022regularizability}.
\end{remark}

\subsubsection{The Learning Phase}
We refer to the main loop of the algorithm in \Cref{line:learning-phase} as the \textit{learning phase}.
During the learning phase, we include temporary ``auxiliary'' links to $\mathcal{G}_d$ and make the communication graph a clique.
We show such distinction by $\mathcal{G}_{d,\mathrm{learn}}$, where $\left\vert \mathcal{V}_{\mathcal{G}_d} \right\vert = \left\vert \mathcal{V}_{\mathcal{G}_{d, \mathrm{learn}}} \right\vert$ but $\mathcal{G}_{d,\mathrm{learn}}$ is a clique.
Inherent to \ac{D2SPI} is a policy iteration on $\mathcal{G}_{d,\mathrm{learn}}$ that characterizes  components $K_k$ and $L_k$, intuitively representing ``self'' and ``cooperative'' controls at iteration $k$, respectively.
In particular, during the learning phase, we utilize these control components in order to design and update an effective stabilizing controller for the rest of the network $\mathcal{G} \setminus \mathcal{G}_{d,\mathrm{learn}}$.

\setlength{\textfloatsep}{0pt}
\begin{algorithm}[!ht]
	\caption{\small \acf{D2SPI}}
	\label{alg:distributed_control_algorithm}
	\begin{algorithmic}[1]
	\small
		\State \textbf{Initialization} $(t \leftarrow 1,\ k \leftarrow 1)$
		
		\State $\hspace{1mm}$ Choose $\mathcal{G}_d \subseteq \mathcal{G}$ with $d = d_{\max}(\mathcal{G})+1$
		
		\State $\hspace{1mm}$ Obtain $Q_1$, $Q_2$, $R$, and set $Q_d \leftarrow Q_1 + d Q_2$
		
		\State $\hspace{1mm}$ Get $\tilde{\mathbf{x}}_1 \in \mathbb{R}^{d n}$ and $\widetilde{\mathbf{H}}_0\in\mathbb{R}^{p\times p}$, $p=d(n+m)$
		
		\State $\hspace{1mm}$ Set
		$\mathcal{P}_0 \leftarrow \beta \mathrm{I}_{p(p+1)/2}$ for large $\beta>0$ and fix $\Sigma$
		
		\State \label{line:assumption} $\hspace{1mm}$Set $K_1$ stabilizing \eqref{eq:agent_i_dynamics}, $L_1 = \mathbf{0}$, $\Delta K_1 = K_1$
		
		\State $\hspace{1mm}$ Turn on temporary links in $\mathcal{G}_d$ and set $\tau_1 \leftarrow 0$
		
		\State \label{line:learning-phase} \textbf{While $(K_k,L_k)$ has not converged, do} (``learning phase'')
				
		\State \label{line:policy_update} $\hspace{1mm}$Set $\mathrm{Policy}_k(\mathcal{V}_{\mathcal{G}})$ such that for each $i\in\mathcal{V}_{\mathcal{G}}$,
		\begin{gather*}
		\begin{aligned}
		    \textstyle \u_{i} &\textstyle \leftarrow \Delta K_k \x_{i} + L_k\sum_{ j\in\mathcal{N}_i} \frac{\tau_k}{d-1}  \x_{j},   &\text{ if } &  i\in\mathcal{V}_{\mathcal{G} \setminus \mathcal{G}_d}\\
		    \textstyle \u_{i} & \textstyle \leftarrow \Delta K_k \x_{i} + L_k\sum_{j \in \mathcal{V}_{\mathcal{G}_d}}  \x_{j},  &\text{ if } &  i\in\mathcal{V}_{\mathcal{G}_d}
		\end{aligned}
		\end{gather*}
        
		\State 	$\hspace{1mm}$ Evaluate $\widetilde{\mathbf{H}}_k $ from \Cref{alg:SPE} \newline
		\small
		$\quad\widetilde{\mathbf{H}}_k \leftarrow \mathrm{SPE} \big( \mathcal{G},\ \mathcal{G}_{d,\mathrm{learn}},\  \mathrm{Policy}_k(\mathcal{V}_{\mathcal{G}}),\ \widetilde{\mathbf{H}}_{k-1}, \mathcal{P}_0 \big)$
        
        \State \label{step:blocks} $\hspace{1mm}$ Recover $X_1,X_2,Y_1,Y_2, Z_1$ and $Z_2$ from $\widetilde{\mathbf{H}}_k$
        \begin{gather*}
        \begin{aligned}
            X_1 &\leftarrow \widetilde{\mathbf{H}}_k[1:n,\ 1:n]\\
            Y_1 &\leftarrow \widetilde{\mathbf{H}}_k[dn+1:dn+m,\ dn+1:dn+m]\\
            Z_1 &\leftarrow \widetilde{\mathbf{H}}_k[dn+1:dn+m,\ 1:n]\\
            X_2 &\leftarrow \widetilde{\mathbf{H}}_k\big[ 1:n,\  n+1:2n \big]\\
            Y_2 &\leftarrow \widetilde{\mathbf{H}}_k[dn+1:dn+m,\ dn+m+1:dn+2m+1]\\
            Z_2 &\leftarrow \widetilde{\mathbf{H}}_k[dn+1:dn+m,\ n+1:2n]\\
            \Delta X &\leftarrow X_1 - X_2, \quad \Delta Y \leftarrow Y_1 - Y_2, \quad \Delta Z \leftarrow Z_1 - Z_2.
        \end{aligned}
        \end{gather*}
        
        \State \label{step:inverses} $\hspace{1mm}$ Update the control components
        \begin{gather*}
        \begin{aligned}
            F^{-1} &\leftarrow Y_1 - (d-1) Y_2 \big( Y_1 + (d-2) Y_2 \big)^{-1} Y_2\\
            G &\leftarrow \left( Y_1 + (d-1) Y_2 \right)^{-1} Y_2 \left( Y_1 - Y_2 \right)^{-1}\\
            K_{k+1} & \leftarrow -F Z_1 + (d-1) G Z_2,\\
            L_{k+1} &\leftarrow -F Z_2 + G Z_1 + (d-2) G Z_2,\\
            \Delta K_{k+1} &\leftarrow K_{k+1} - L_{k+1}
        \end{aligned}
        \end{gather*}

        \State $\hspace{1mm}$ \label{step:singular_values} Obtain the stability margin
        \begin{gather*}
        \begin{aligned}
             \Xi_{k+1} \leftarrow& \Delta X - Q_d + \Delta K_{k+1}^\intercal \Delta Z \\
            &+ \Delta Z^\intercal \Delta K_{k+1} + \Delta K_{k+1}^\intercal \big( \Delta Y - R \big) \Delta K_{k+1}\\
            \gamma_{k+1} \leftarrow& \lambda_{\min} \Big( \Delta K_{k+1}^\intercal R \Delta K_{k+1} + Q_d \Big) \\
            &\big/ \lambda_{\max} \Big( \Xi_{k+1} + L_{k+1}^\intercal (\Delta Y - R) L_{k+1} \Big)\\
            \tau_{k+1} \leftarrow& \sqrt{\gamma_{k+1}^2 / (1 + \gamma_{k+1})}
        \end{aligned}
        \end{gather*}
        \State $\hspace{1mm}$ Go to \Cref{line:learning-phase} and set $k \leftarrow k + 1$
        
        \State $\hspace{0mm}$Switch OFF the temporary links and retrieve $\mathcal{G}_d$
        
        \State \label{step:final_design} $\hspace{0mm}$\label{step:final-policy}Set $\mathrm{Policy}_k(\mathcal{V}_{\mathcal{G}})$ such that for each $i\in\mathcal{V}_{\mathcal{G}}$, \newline
        $\textstyle\u_{i} \leftarrow \Delta K_k \x_{i} + \frac{\tau_k}{d-1} L_k \sum_{j\in\mathcal{N}_i} \x_{j}$
	\end{algorithmic}
\end{algorithm}

We do so by ensuring that during the learning phase, information is exchanged uni-directionally from $\mathcal{G}_{d,\mathrm{learn}}$ to the rest of the network; hence, the policy of the agents in $\mathcal{G} \setminus \mathcal{G}_{d,\mathrm{learn}}$ is dependent on those in $\mathcal{G}_{d,\mathrm{learn}}$, and not vice versa.
After the learning phase terminates,
we remove the temporary links added during
the learning phase (re-initialize
to the original network), and synthesize a
suboptimal stabilizing control for the entire network.
In the learning phase of \ac{D2SPI}, we use a \ac{RLS}-based recursion to estimate the unknown parameters of the $\mathcal{Q}$-function at iteration $k$, referred to as $\widetilde{\mathbf{H}}_k$.

\subsubsection{The \acf{SPE} Subroutine}
This process is performed in \ac{SPE} (\Cref{alg:SPE}) subroutine by inputting (sub-)graph $\mathcal{G}$, $\mathcal{G}_{d,\text{learn}}$, the mapping $\mathrm{policy}(\mathcal{V}_{\mathcal{G}_d})$, and the previous estimate of $\widetilde{\mathbf{H}}_{k-1}$.
As will be discussed in \Cref{subsec:analysis_setup}, $\widetilde{\mathbf{H}}_{k}$ contains the required information to determine the two control components $K_k$ and $L_k$
from data.
We extract this square matrix through a recursive update on the vector $\otheta_{k-1}$, derived from half-vectorization of $\widetilde{\mathbf{H}}_{k-1}$, solving \ac{RLS} for the linear equation $\mathcal{R}(\tilde{\mathbf{x}}_t, \tilde{\mathbf{u}}_t) = \ozeta_t^{\intercal} \otheta_{k-1}$, where $\mathcal{R}(\tilde{\mathbf{x}}_t, \tilde{\mathbf{u}}_t)$ denotes the local cost and $\ozeta_t \in \mathbb{R}^p$ contains the data measurements.
{We use subscript $k$ for policy update and $t$ for data collection.}
The adaptive nature of the algorithm involves
the exploration signal $\e_t$ to be augmented to the policy vector in order to provide persistence of excitation.

\subsubsection{Persistence of Excitation and Convergence of \ac{SPE}}
In our setup, $\e_t$ is sampled from a normal distribution $\e \sim \mathcal{N}(\mathbf{0},\Sigma)$ where the choice of the variance $\Sigma \succ 0$ is problem-specific.
{In practice, excitation of the input is a subtle task and has been realized in a variety of forms such as random noise \cite{bradtke1994adaptive}, sinusoidal signals \cite{jiang2012computational}, and exponentially decaying noise \cite{lewis2010reinforcement}.}
We denote by $\mathcal{P}$ the projection factor that is reset to $\mathcal{P}_0 \succ 0$ for each iteration.
Convergence of \ac{SPE}---guaranteed based on the persistence of excitation condition---is followed by the update of $\widetilde{\mathbf{H}}_k$ that encodes the necessary information to obtain $K_k$ and $L_k$.

\subsubsection{Learning Control Components using Patterned Monoid}
Learning the control components $K_k$ and $L_k$ is achieved by first recovering the block matrices $X_1$, $X_2$, $Y_1$, $Y_2$, $Z_1$, and $Z_2$ from $\widetilde{\mathbf{H}}_k$ that are further utilized to form intermediate variables $F$ and $G$. Matrix inversions on \cref{step:inverses} of \Cref{alg:distributed_control_algorithm} will be justified in \Cref{sec:analysis} \Cref{lem:tools_for_analysis}.
Such recovery of meaningful blocks from $\widetilde{\mathbf{H}}_k$ is due to the specific structure resulting from adding extra links to $\mathcal{G}_d$ that is captured systematically by the Patterned monoid; this
point will be discussed further subsequently in \Cref{prop:H_structure}.

\subsubsection{Learning Gain Margin and the Distributed Feedback Design}
Each iteration loop is completed by updating the parameters $\gamma_k$ and $\tau_k$,
that prove instrumental in the stability analysis of the proposed distributed controller for the entire network.
Finally, with the convergence of \ac{D2SPI}, $\mathcal{G}_d$ is retrieved by removing the temporary links and the structured policy is extended to the entire graph $\mathcal{G}$ in \Cref{step:final_design}.

Let us point out a few remarks on the computational
complexity of the proposed algorithm.
First, note that the inverse operations on \cref{step:inverses} occur on matrices of size $m\times m$,
and hence computationally inexpensive.
Furthermore, the complexity of finding extreme singular values---as on \cref{step:singular_values} in \Cref{alg:distributed_control_algorithm}---is known
to be $\mathcal{O}(n^2)$ \cite{comon1990tracking}.
Hence, the computational complexity of \ac{D2SPI} is mainly due to the \ac{SPE} recursion that is equivalent to the complexity of \ac{RLS} for the number of unknown system parameters in $\mathcal{G}_d$, i.e., the computational
cost is $\mathcal{O} \left( d^2 (n+m)^2 \right)$ \cite{haykin2002adaptive}.
This implies that the computational complexity of the algorithm is {\em fixed} for any number of agents $N$, as long as the maximum degree of the graph 
retains its order.
\begin{remark}
    \label{rmk:notes_on_algorithm}
    
    Adding temporary links within the subgraph $\mathcal{G}_d$ is an effective means of learning optimal $K_k$ and $L_k$ for the subgraph $\mathcal{G}_{d,\mathrm{learn}}$ by utilizing dynamical interdependencies among the agents.
    Although initializing $K_k$ such that \eqref{eq:agent_i_dynamics} is Schur stable is a standard assumption in data-driven control, obtaining this initial gain for an unknown system is nontrivial.
    While we invoke this assumption in this work, the interested reader is referred to~\cite{talebi2020online, chen2020black} for more recent works pertaining to this
    assumption and related system theoretic issues~\cite{yue2021controllability}.
\end{remark}


\section{Convergence and Stability Analyses}
\label{sec:analysis}

In this section, we provide convergence and stability analyses for the \ac{D2SPI} algorithm.
In this direction, we first study the structure and stability  margins of each local controller and proceed to establish  stability properties  of the proposed controller for the entire network throughout the learning process.
%
Lastly, we show the convergence of \ac{D2SPI} to a stabilizing suboptimal distributed controller followed by the derivation of a suboptimality bound characterized by the problem parameters. 
For clarity, we defer some of the subtleties 
of the analysis and detailed proofs to Appendix \ref{sec:setup-analysis}.

First, let us demonstrate how a specific structure and stability of the controller for the subgraph $\mathcal{G}_{d,\mathrm{learn}}$, when properly initialized, can be preserved throughout the \ac{D2SPI} algorithm.
 
 \begin{proposition}
    \label{prop:H_structure}
   Let $\widetilde{\mathbf{K}}_k \coloneqq \mathrm{I}_d \otimes \big( K_k - L_k \big) + \mathbbm{1} \mathbbm{1}^{\intercal} \otimes L_k$ for all $k \geq 1$, with $K_k$ and $L_k$ as in \Cref{alg:distributed_control_algorithm}.
    Under \Cref{assmp:main} and throughout the learning phase (for all $k \geq 1$), $\widetilde{\mathbf{K}}_k$ is stabilizing for the system in $\mathcal{G}_{d,\mathrm{learn}}$ and $\mathrm{Policy}_k( \mathcal{V}_{\mathcal{G}_{d,\text{learn}}}) |_t = \widetilde{\mathbf{K}}_k \, \mathrm{State} (\mathcal{V}_{\mathcal{G}_{d,\text{learn}}})|_t,$ for all $t$. Furthermore, $\Delta K_k \coloneqq K_k - L_k$ stabilizes the dynamics of a single agent, i.e., $A+B \Delta K_k$ is Schur stable.
\end{proposition}

Note that \Cref{prop:H_structure} proves the existence of a stabilizing controller $\Delta K_k$ and its corresponding cost-to-go matrix $\Delta P_k$.
In the sequel, our goal is to design a distributed suboptimal controller for the entire networked system on $\mathcal{G}$ based on the components that shape $\Delta K_k$. This
extension is built upon the stability margin 
derived next.

\begin{proposition}
    \label{prop:gain_margin}
        At each iteration $k \geq 1$, let $K_k$, $L_k$ and $\tau_k$ be obtained via \Cref{alg:distributed_control_algorithm}.
    Then, $A + B(K_k - \alpha L_k)$ is Schur stable for all $\alpha$ satisfying $\left\vert \alpha - 1 \right\vert \leq \tau_k$.
\end{proposition}

The stability margin $\tau_k$ in
\Cref{prop:gain_margin} is upper-bounded by the stability margin of the pair $(A + B(K_k-L_k), B)$. 
This implies that if the original closed-loop system for an  agent does not have a favorable stability margin, then $\tau_k$ can be small--reducing the influence of the agent's neighbors on its policy (\Cref{step:final-policy} of \Cref{alg:distributed_control_algorithm}).
Nonetheless, \Cref{prop:gain_margin} provides model-free stability gain margins $\tau_k$ at each iteration of the algorithm for the dynamics of a single agent in $\mathcal{G}$.
In our analysis, we take advantage of these margins to 
characterize stability guarantees for the controller proposed during the learning phase of \ac{D2SPI}. This is captured in the following result.

\begin{theorem}
    \label{thm:stabilizability_learning_phase}
    Suppose $K_k$, $L_k$ and $\tau_k$ are defined as in \Cref{alg:distributed_control_algorithm}.
    Then, under \Cref{assmp:main}, the control policy $\mathrm{Policy}_k(\mathcal{V}_{\mathcal{G}})$ designed during the learning phase (\cref{line:policy_update})
    stabilizes the network $\mathcal{G}$ at each iteration of the learning phase and for any choice of $\mathcal{V}_{\mathcal{G}_d}$.
\end{theorem}

\Cref{thm:stabilizability_learning_phase}
establishes that the proposed feedback mechanism stabilizes the entire network, facilitating control of agents {\em outside} of $\mathcal{G}_d$, during the learning phase.
In the meantime, the practicality and suboptimality of the algorithm depend on its convergence addressed next.
\begin{theorem}
    \label{thm:convergence}
    
    Under \Cref{assmp:main} and (long enough) finite termination of \Cref{alg:SPE}, \Cref{alg:distributed_control_algorithm} converges, i.e., $K_k \rightarrow K^*$, $L_k \rightarrow L^*$, and $\tau_k \rightarrow \tau^*$ as $k\rightarrow\infty$, where $\widetilde{\mathbf{K}}^* = \mathrm{I}_d \otimes \big( K^* - L^* \big) + \mathbbm{1} \mathbbm{1}^{\intercal} \otimes L^*$ is the optimal solution to the infinite-horizon state-feedback LQR problem with system parameters $\big( \widetilde{\mathbf{A}}, \widetilde{\mathbf{B}}, \widetilde{\mathbf{Q}}, \widetilde{\mathbf{R}} \big)$ defined as $ \widetilde{\mathbf{A}} = \mathrm{I}_d \otimes A$, $\widetilde{\mathbf{B}} = \mathrm{I}_d \otimes B$, $\widetilde{\mathbf{Q}} = \mathrm{I}_{d} \otimes (Q_1 + d Q_2) - \mathbbm{1}\mathbbm{1}^{\intercal} \otimes Q_2$, and $\widetilde{\mathbf{R}} = \mathrm{I}_d \otimes R$.
\end{theorem}

Finally, we note that as the temporary links 
introduced during the learning phase are removed, the structure of the agents' interaction is once again the original
network $\mathcal{G}$.
As such, it is vital to provide stability guarantees after \Cref{alg:distributed_control_algorithm} terminates and components of the control design have
converged.
This issue is addressed in the following corollary whose proof is similar to that of \Cref{thm:stabilizability_learning_phase} and thus omitted for brevity.

\begin{algorithm}[!ht]
	\caption{\small {Subgraph Policy Evaluation} (SPE)}
	\label{alg:SPE}
	\begin{algorithmic}[1]
	\small
	\State $\hspace{0mm}$ \textbf{Input:} Graph $\mathcal{G}$, subgraph $\mathcal{G}'\subseteq \mathcal{G}$, $\mathrm{Policy}(\mathcal{V}_{\mathcal{G}})$, $\mathbf{H}$, $\mathcal{P}$

	 \State $\hspace{0mm}$ \textbf{Output:} Updated cost matrix $\mathbf{H}^+$ associated with $\mathcal{G}'$

	\State $\hspace{0mm}$ \textbf{While} $\mathbf{H}$ \textbf{has not converged, do}

	\State $\hspace{2mm}$
	Set $\tilde{\mathbf{x}}_t \leftarrow \mathrm{State} (\mathcal{V}_{\mathcal{G}'})|_t$ and $\tilde{\mathbf{u}}_t \leftarrow \mathrm{Policy} (\mathcal{V}_{\mathcal{G}'})|_t$

	\State $\hspace{2mm}$ Choose $\e_t \sim \mathcal{N}(\mathbf{0},\Sigma)$ and update $\mathrm{Policy}(\mathcal{V}_{\mathcal{G}'}) |_t$ as $\tilde{\mathbf{u}}_t \leftarrow \tilde{\mathbf{u}}_t + \e_t$  for all $i\in \mathcal{V}_{\mathcal{G}'}$

	\State $\hspace{2mm}$ Run $\mathcal{G}$ under policy $\mathrm{Policy}(\mathcal{V}_{\mathcal{G}})$

	\State $\hspace{2mm}$ Collect $\mathrm{State} (\mathcal{V}_{\mathcal{G}'})|_{t+1}$ only from $\mathcal{G}'$ and set \newline
	$\qquad \tilde{\mathbf{x}}_{t+1} \leftarrow \mathrm{State} (\mathcal{V}_{\mathcal{G}'})|_{t+1}, \; \tilde{\mathbf{u}}_{t+1} \leftarrow \mathrm{Policy}( \mathcal{V}_{\mathcal{G}'}) |_{t+1}$

    \State $\hspace{2mm}$ Set $\ophi_t \leftarrow [\tilde{\mathbf{x}}_{t}^\intercal ~ \tilde{\mathbf{u}}_{t}^\intercal]^\intercal - [\tilde{\mathbf{x}}_{t+1}^\intercal ~ \tilde{\mathbf{u}}_{t+1}^\intercal]^\intercal$

	\State $\hspace{2mm}$ Compute $\ozeta_t = \mathrm{vech}(\ophi_t \ophi_t^\intercal)$ and \newline
    \small
	$\mathcal{R}(\tilde{\mathbf{x}}_t, \tilde{\mathbf{u}}_t) = \tilde{\mathbf{x}}_t^{\intercal} \big( \mathrm{I} \otimes Q_d - \mathbbm{1}\mathbbm{1}^{\intercal} \otimes Q_2 \big) \tilde{\mathbf{x}}_t + \tilde{\mathbf{u}}_t^{\intercal} \big( \mathrm{I} \otimes R \big) \tilde{\mathbf{u}}_t$

	\State $\hspace{2mm}$ \label{eq:RLS_iteration}Set $\otheta \leftarrow \mathrm{vech}(\mathbf{H})$ and update \newline
	\small
	$\begin{aligned}
	    \otheta \leftarrow& \otheta + {\mathcal{P} \ozeta_{t} \big( \mathcal{R}(\tilde{\mathbf{x}}_t, \tilde{\mathbf{u}}_t) - \ozeta_{t}^\intercal \otheta \big)}\big/{(1 + \ozeta_{t}^\intercal \mathcal{P} \ozeta_{t})},\\
	    \mathcal{P} \leftarrow& \mathcal{P} - {\mathcal{P} \ozeta_{t} \ozeta_{t}^\intercal \mathcal{P}}\big/{(1 + \ozeta_{t}^\intercal \mathcal{P} \ozeta_{t})}
	\end{aligned}$

	\State $\hspace{2mm}$ Find $\mathbf{H}^+ = \mathrm{vech}^{-1}(\otheta)$, update $\mathbf{H} \leftarrow \mathbf{H}^+$ and $t\leftarrow t+1$


	\end{algorithmic}
\end{algorithm}

\begin{corollary}
    \label{cor:ultimate_stabilizability}
    
    Suppose that $K^*$, $L^*$, $\gamma^*$, and $\tau^*$ are given as in \Cref{thm:convergence} under a convergent \Cref{alg:distributed_control_algorithm}.
    Then $\mathrm{Policy}(\mathcal{V}_{\mathcal{G}})$ (defined on \Cref{step:final-policy}),
    stabilizes the entire networked system in \eqref{eq:dynamics_N_agents}.
\end{corollary}

We conclude this section by exploring the suboptimality of the proposed policy.
Given the problem parameters, let $\hat{\mathbf{K}}_{\mathrm{struc}}^*$ denote the globally optimal distributed solution for the structured LQR problem in \eqref{eq:distributed_optimization} with the associated cost matrix $\hat{\mathbf{P}}_{\mathrm{struc}}^*$.
Given any other stabilizing structured policy $\hat{\mathbf{K}}$ associated with cost matrix $\hat{\mathbf{P}}$, we define the \textit{optimality gap} as
\(\mathrm{\text{gap}}(\hat{\mathbf{K}}) \coloneqq \mathrm{Tr}[\hat{\mathbf{P}} - \hat{\mathbf{P}}_{\mathrm{struc}}^*].\)
 The following theorem provides an upperbound on the optimality gap of structured policy learned by \ac{D2SPI} based on the problem parameters. In particular, when the system is ``contractible,'' the derived upperbound depends on the difference of the distributed controller with that of {\em unstructured} optimal LQR controller.

\begin{theorem}
    \label{thm:suboptimality}
    Let $\hat{\mathbf{K}}^*$ be the structured policy learned by \Cref{alg:distributed_control_algorithm} at convergence, corresponding to the cost matrix $\hat{\mathbf{P}}^*$.
    Moreover, let $\hat{\mathbf{K}}_{\mathrm{lqr}}$ denote the optimal (unstructured) solution to the infinite-horizon state-feedback LQR problem with parameters
    $(\hat{\mathbf{A}}, \hat{\mathbf{B}}, \hat{\mathbf{Q}}, \hat{\mathbf{R}})$ with the cost matrix $\hat{\mathbf{P}}_{\mathrm{lqr}}$.
    If $\hat{\mathbf{A}}_{\hat{\mathbf{K}}_{\mathrm{lqr}}} = \hat{\mathbf{A}} + \hat{\mathbf{B}} \hat{\mathbf{K}}_{\mathrm{lqr}}$ is contractible then
    \(
        0 \leq \mathrm{gap}(\hat{\mathbf{K}}^*) \leq {\mathrm{Tr}(\mathbf{M})}/{[1-\sigma_{\max}^2(\hat{\mathbf{A}}_{\hat{\mathbf{K}}_{\mathrm{lqr}}})]},
    \)
    where
   $\mathbf{M} \coloneqq (\hat{\mathbf{R}}+ \hat{\mathbf{B}}^\intercal \hat{\mathbf{P}}^* \hat{\mathbf{B}})  ( \hat{\mathbf{K}}^*  \hat{\mathbf{K}}^{*\intercal} -  \hat{\mathbf{K}}_{\mathrm{lqr}} \hat{\mathbf{K}}_{\mathrm{lqr}}^\intercal) + 2 \hat{\mathbf{A}}^\intercal \hat{\mathbf{P}}^*  \hat{\mathbf{B}} (\hat{\mathbf{K}}^* - \hat{\mathbf{K}}_{\mathrm{lqr}}).$
\end{theorem}

\begin{remark}
    \label{rmk:contractibility}
    
    First, recall how the converged policy by \Cref{thm:convergence} is related to the optimal \ac{LQR} policy on the fully connected subgraph $\mathcal{G}_{d,\mathrm{learm}}$. 
    Second, note that the optimality gap is characterized by ${\mathrm{Tr}(\mathbf{M})}$ which is essentially proportional to the difference $\hat{\mathbf{K}}^* -  \hat{\mathbf{K}}_{\mathrm{lqr}}$; i.e., how close our designed structured policy is to the unstructured LQR that directly depends on the connectivity of the graph topology.
    Third, the contractibility of the pair $(A,B)$ is more restrictive condition than regularizability of the system \cite{talebi2020online}, a notion that has recently been employed in iterative data-guided control methods \cite{lale2020regret, agarwal2019online}. 
    Contractibility also facilitates the validity of assuming access to the initial stabilizing controller.
\end{remark}


\section{Simulation Examples}
\label{sec:simulation}
In this section, we examine the performance and convergence of
\ac{D2SPI}.
In order to assess the suboptimality of the synthesized
controllers, we report the trace of cost matrices, $\mathrm{Tr}(\hat{\mathbf{P}}_k)$, associated with the proposed distributed controller learned by \ac{D2SPI} at iteration $k$. As the optimal distributed design is unknown, we compare these results against the optimal cost for the unconstrained LQR problem, $\mathrm{Tr}(\hat{\mathbf{P}}_{\text{LQR}})$, obtained
via the solution of the Algebraic Riccati Equation with parameters $(\hat{\mathbf{A}}, \hat{\mathbf{B}}, \hat{\mathbf{Q}}, \hat{\mathbf{R}})$. Note that this is an infeasible solution to the problem in \eqref{eq:distributed_optimization}; nevertheless, it provides a theoretical lowerbound to evaluate the performance of any feasible solution---including the optimal one.\footnote{All the simulations were run on a 3.2 GHz Quad-Core Intel Core i5 CPU and in MATLAB. The scripts are publicly available at \url{https://github.com/shahriarta/D2SPI}.}

\subsection{Convergence--Randomly Selected Parameters}\label{exp:random-paramters}
In the first example, we sample continuous-time system parameters of a single agent $(A,B)$ from a zero-mean normal distribution with unit covariance, 
such that $A \in \mathbb{R}^{5\times 5}$ and $B \in \mathbb{R}^{5\times 3}$. 
We then consider a path-graph of 10 agents and demonstrate how \Cref{alg:distributed_control_algorithm} converges for this network using different instances of the system parameters.
The continuous-time system dynamics of a single agent is discretized with a sampling rate of $\Delta T = 0.1s$.
We set $Q_1 = 0.2$, $Q_2 = \mathrm{I}_n$ and $R = \mathrm{I}_m$ in \cref{eq:distributed_optimization}.
We assume a random exploration signal sampled from a normal distribution $e_t\sim\mathcal{N}(0,\sigma^2)$, where the variance $\sigma^2$ is chosen accordingly for different input channels.

\begin{figure}[t]
    \centering
    \includegraphics[width=0.9\columnwidth]{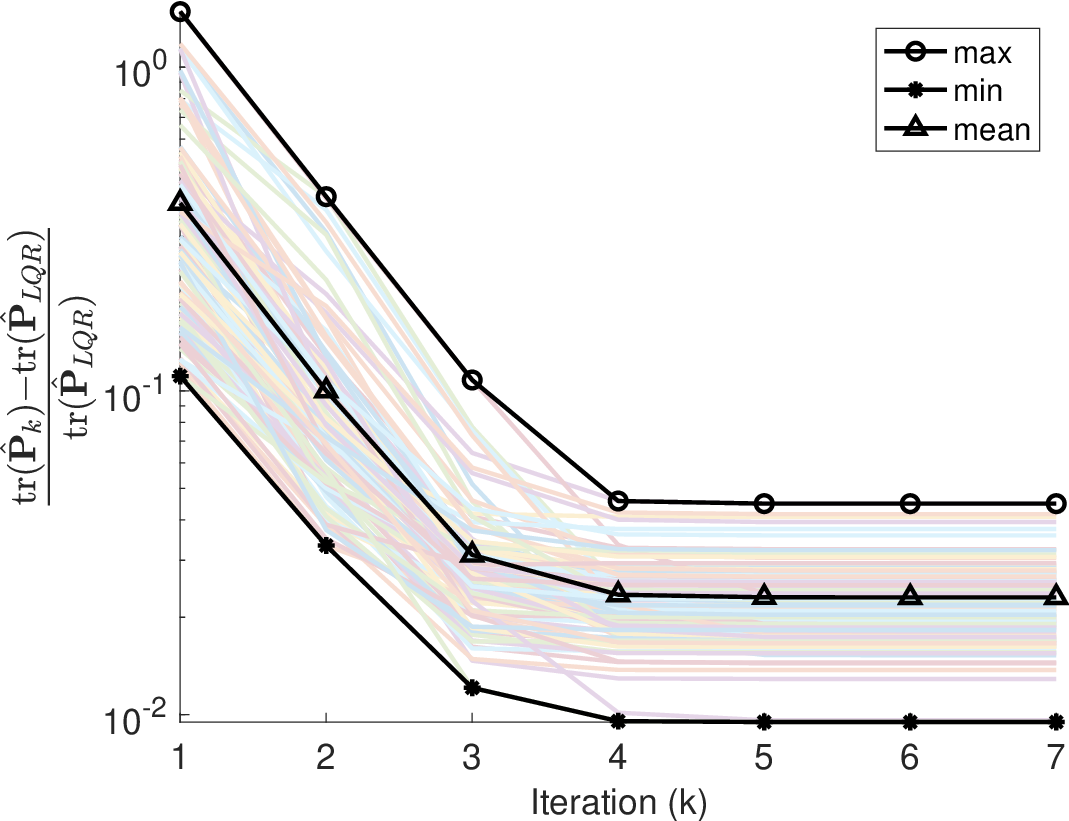}
    \caption{\small Convergence of \Cref{alg:distributed_control_algorithm} for 100 randomly sampled system parameters $(A,B)$.}
    \label{fig:exp1}
\end{figure}

\Cref{fig:exp1} shows the performance of the synthesized controller by illustrating the normalized suboptimality error  for the entire network with respect to the (infeasible) LQR controller.
This figure depicts the simulation results for 100 random system parameters and shows the progress of the proposed method at each iteration in \Cref{alg:distributed_control_algorithm}.
The actual simulations are plotted in faded color, whereas their statistical characteristics are plotted in solid black.
Note that almost all realizations of
the network have converged after 5 iterations of the learning phase, due to its quadratic convergence rate.
Assessing the suboptimality of the proposed controller compared against the (infeasible) centralized LQR controller, reveals an average improvement by a factor of 200.

\subsection{Network of homogeneous plants}
\label{sec:simulation2}
In this example, we apply \ac{D2SPI} to two other 
simulation scenarios involving homogeneous networks of agents with unknown and unstructured model uncertainties.
In particular, we use the dynamics of plants with continuous-time system parameters $(A,B)$ (as reported in Appendix F of \cite{hung1982multivariable}),
in conjunction with random $d$-regular graph topologies of different sizes.
We then examine the efficacy of \ac{D2SPI} by illustrating the cost associated with the proposed distributed controller as a function of nodes in the graph.
The rest of the problem parameters are chosen identical to the setup in \Cref{exp:random-paramters}. 
Uncertainty in the model are introduced in this
example as follows.
Each agent follows an unknown LTI dynamics similar to \eqref{eq:agent_i_dynamics} with $A$ replaced by $ A + \Delta A $,
where entries of $\Delta A$ are sampled from a normal distribution $\mathcal{N}(0, 0.05)$. 
Assuming that one has access to a stabilizing controller $K_1$ for the system with nominal parameters $A$ and $B$, we set the initial stabilizing controller to be the LQR optimal controller with parameters $(A,B,Q_1,R)$.
\begin{figure*}
    \centering
    \begin{subfigure}{0.49\columnwidth}
        \centering
        \includegraphics[width=\columnwidth]{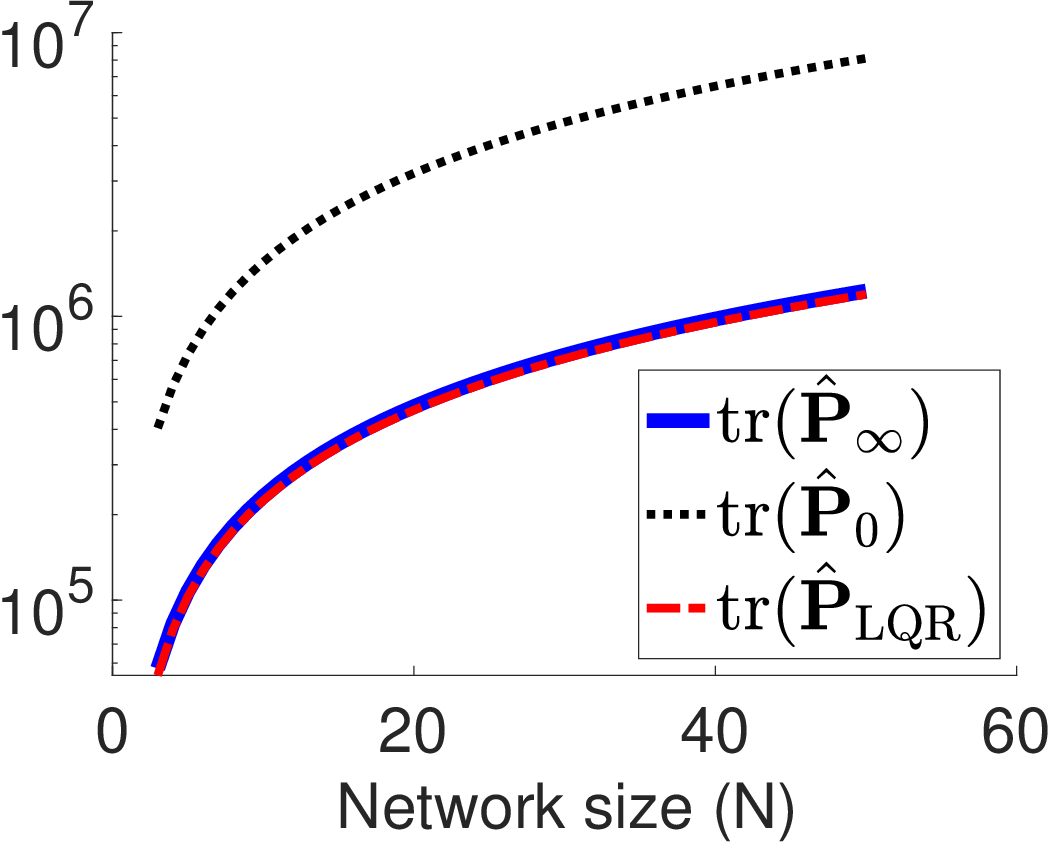}
        \caption{}
        \label{fig:exp2-comparison}
    \end{subfigure}
    \hfill
    \begin{subfigure}{0.49\columnwidth}
        \centering
        \includegraphics[width=\columnwidth]{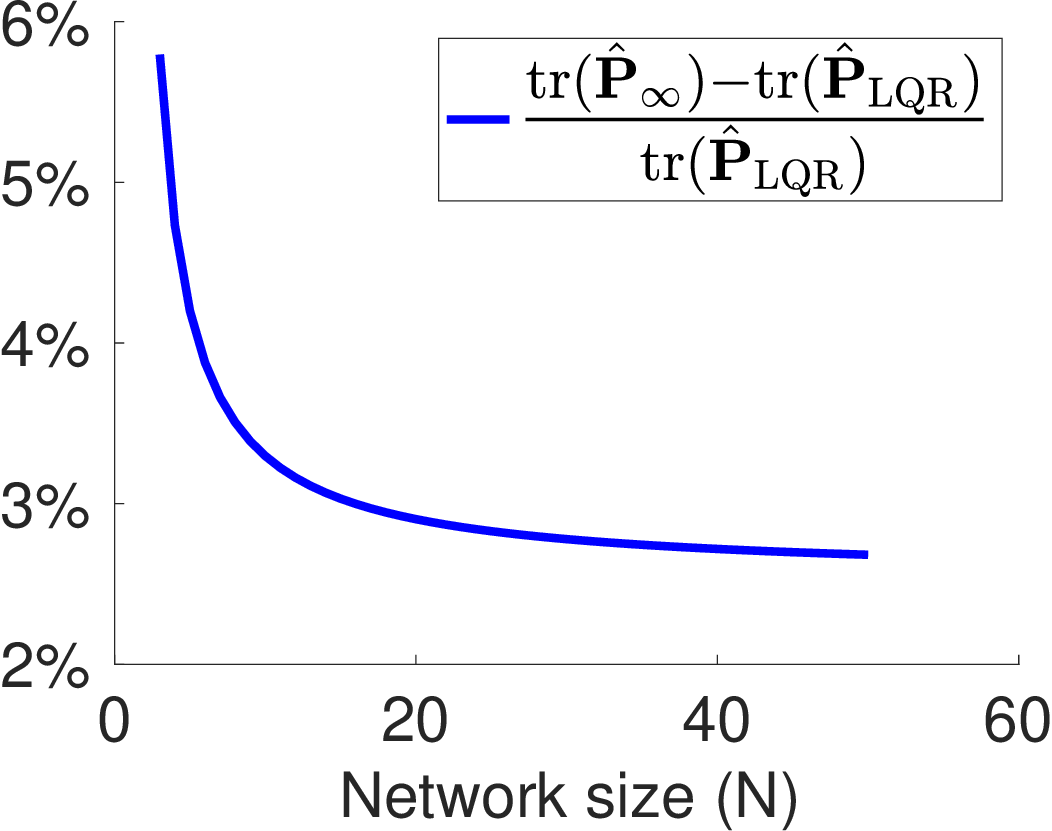}
        \caption{}
        \label{fig:exp2-error}
    \end{subfigure}
    \begin{subfigure}{0.49\columnwidth}
        \centering
        \includegraphics[width=\columnwidth]{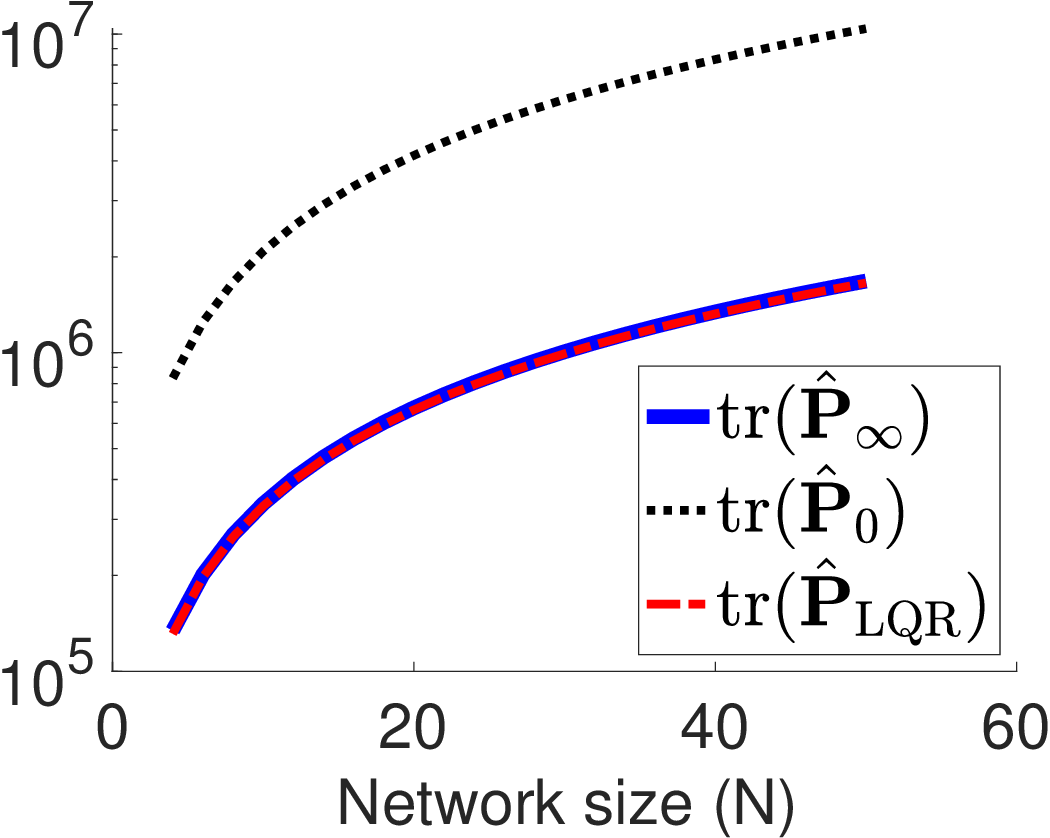}
        \caption{}
        \label{fig:exp3-comparison}
    \end{subfigure}
    \hfill
    \begin{subfigure}{0.49\columnwidth}
        \centering
        \includegraphics[width=\columnwidth]{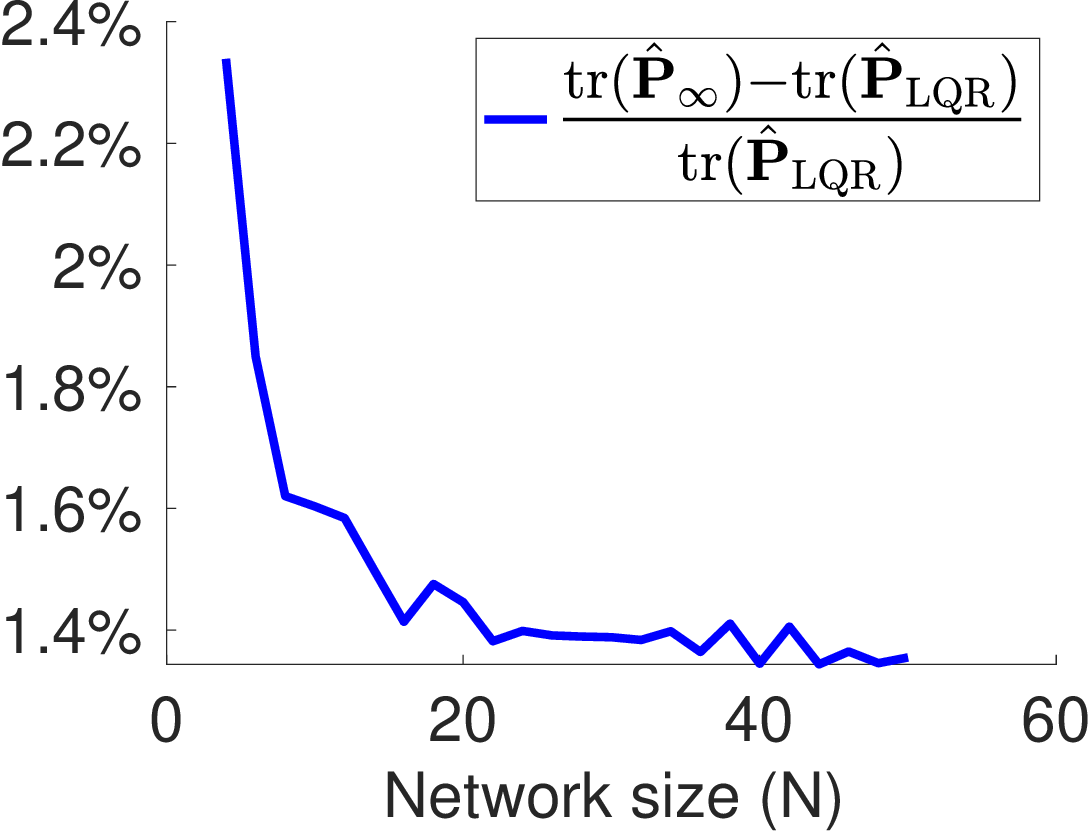}
        \caption{}
        \label{fig:exp3-error}
    \end{subfigure}
    \caption{ Suboptimality of the distributed controller learned by \Cref{alg:distributed_control_algorithm} for a network of $N$ homogeneous plants with path-graphs structures in (a) and (b), and with random 3-regular graphs structures in (c) and (d).}
    \label{fig:exp2-3}
    \vspace{-0.5cm}
\end{figure*}

\Cref{fig:exp2-3} shows the results of the second simulation example, illustrating how the cost of the proposed controller changes  with respect to the number of nodes in a path graph  with different number of nodes.
\Cref{fig:exp2-comparison} compares the cost associated with our design $\hat{\mathbf{P}}_{\infty}$ against the cost of the initial controller $\hat{\mathbf{P}}_1$, and the (infeasible) LQR controller $\hat{\mathbf{P}}_{\text{LQR}}$.
\Cref{fig:exp2-error} illustrates the evolution of the normalized suboptimality of our proposed algorithm (with respect to the infeasible LQR controller) as a function of number of nodes in the corresponding graphs. 
\Cref{fig:exp3-comparison} and \Cref{fig:exp3-error} show similar results for the random 3-regular graph 
topologies with even number of nodes.

As it can be validated from \Cref{fig:exp2-3}, the cost associated with our final proposed distributed controller has significantly improved the optimality of the initial controller. In particular, the normalized suboptimality errors of our final design are less than 6\% and 2.4\% for path-graph and random 3-regular graph topologies, respectively. Furthermore, this normalized error
generally decrease as the number of nodes in the 
corresponding graphs increase.

\section{Concluding Remarks}
\label{sec:conclusion}

In this paper, we have proposed the \ac{D2SPI} algorithm as an efficient model-free distributed control synthesis
process for potentially high-dimensional network of
homogeneous linear systems.
\ac{D2SPI} is built upon a construction 
referred to as Patterned monoid,
that facilitates exploiting network symmetries
for policy synthesis consistent with the underlying
network.
Such symmetries allow a data collection procedure during the learning phase (with temporary additional links) for a smaller portion of the network.
Using data collection on this smaller subnetwork,
we are then able to synthesize a distributed feedback mechanism for the entire system--even during the learning phase.
Moreover, \ac{D2SPI} builds upon parameter estimation techniques that represent an end-to-end policy prediction directly from the observed data.
Extension of the setup proposed for \ac{D2SPI} 
involves heterogeneous networked systems,
that is currently being examined as part of our future work.

\appendices

\section{Analysis and Proofs}
\label{sec:setup-analysis}

In this section, we provide the building blocks needed for the proofs and analysis of our algorithm.
We first provide some insights on how the setup is connected to the classic model-based LQR machinery and some previously established results that we leverage from the literature. The main proofs then follows.
%

\subsection{Underlying Model of the Subsystem \texorpdfstring{$\mathcal{G}_d$}{Gd}}
\label{subsec:analysis_setup}

The configuration of the synthesis problem in \ac{D2SPI} intertwines an online recursion on the subsystem corresponding to $\mathcal{G}_d$ and the original system $\mathcal{G}$.
In particular, during the learning phase, considering the same cost structure and problem parameters as in \eqref{eq:distributed_optimization}---but for the completed subgraph $\mathcal{G}_{d,\mathrm{learn}}$--- results in the $\big( \widetilde{\mathbf{A}}, \widetilde{\mathbf{B}}, \widetilde{\mathbf{Q}}, \widetilde{\mathbf{R}} \big)$ parameters as defined in \Cref{thm:convergence}.
Then, similarly from \eqref{eq:agent_i_dynamics} the dynamics of the subgraph $\mathcal{G}_{d,\mathrm{learn}}$ assumes the form,
\begin{gather}
    \label{eq:tilded_dynamics}
    \tilde{\mathbf{x}}_{t+1} = \widetilde{\mathbf{A}} \tilde{\mathbf{x}}_{t} + \widetilde{\mathbf{B}} \tilde{\mathbf{u}}_{t},
\end{gather}
where $\tilde{\mathbf{x}}_t$ and $\tilde{\mathbf{u}}_t$ are formed from concatenation of state and control signals in $\mathcal{G}_{d,\mathrm{learn}}$---recall that $\tilde{\mathbf{u}}_t$ is also denoted by $\mathrm{Policy}(\mathcal{V}_{\mathcal{G}_{d,\mathrm{learn}}})|_t$ in the algorithm to emphasize the temporal implementation of a specific policy in \Cref{alg:SPE}.
From the Bellman equation \cite{bellman1966dynamic} for the LQR problem with these parameters, the cost matrix $\widetilde{\mathbf{P}}_k$ of $\mathcal{G}_{d,\mathrm{learn}}$ is correlated with the one-step LQR cost as,
\begin{align}
    \label{eq:bellman_equation}
    \tilde{\mathbf{x}}_t^{\intercal} \widetilde{\mathbf{P}}_k \tilde{\mathbf{x}}_t = \mathcal{R}(\tilde{\mathbf{x}}_t, \tilde{\mathbf{u}}_t) + \tilde{\mathbf{x}}_{t+1}^{\intercal} \widetilde{\mathbf{P}}_k \tilde{\mathbf{x}}_{t+1},
\end{align}
where $\mathcal{R}(\tilde{\mathbf{x}}_t, \tilde{\mathbf{u}}_t) = \tilde{\mathbf{x}}_t^{\intercal} \widetilde{\mathbf{Q}} \tilde{\mathbf{x}}_t + \tilde{\mathbf{u}}_t^{\intercal} \widetilde{\mathbf{R}} \tilde{\mathbf{u}}_t$ and $\widetilde{\mathbf{P}}_k$ satisfies the Lyapunov equation,
\begin{gather}
    \begin{aligned}[b]
    \label{eq:tilded_lyapunov}
    \widetilde{\mathbf{P}}_k = \big( \widetilde{\mathbf{A}} + \widetilde{\mathbf{B}} \widetilde{\mathbf{K}}_k \big)^{\intercal} \widetilde{\mathbf{P}}_k \big( \widetilde{\mathbf{A}} + \widetilde{\mathbf{B}} \widetilde{\mathbf{K}}_k \big) + \widetilde{\mathbf{Q}} + \widetilde{\mathbf{K}}_k^{\intercal} \widetilde{\mathbf{R}} \widetilde{\mathbf{K}}_k,
    \end{aligned}
\end{gather}
and $\widetilde{\mathbf{K}}_k$ is the controller policy at iteration $k$.
The dynamic programming solution to the LQR problem suggests a linear feedback form $\tilde{\mathbf{u}}_t = \widetilde{\mathbf{K}}_k \tilde{\mathbf{x}}_t$ for the subsystem $\mathcal{G}_{d,\mathrm{learn}}$ at each iteration.
Combining \eqref{eq:tilded_dynamics} and \eqref{eq:bellman_equation} with some rearrangements result in
\begin{align}
    \label{eq:H_has_blocks}
    \tilde{\mathbf{x}}_t^{\intercal} \widetilde{\mathbf{P}}_k \tilde{\mathbf{x}}_t &= \tilde{\mathbf{z}}_t^{\intercal} \begin{bmatrix}
        \widetilde{\mathbf{Q}} + \widetilde{\mathbf{A}}^{\intercal} \widetilde{\mathbf{P}}_k \widetilde{\mathbf{A}} & \widetilde{\mathbf{A}}^{\intercal} \widetilde{\mathbf{P}}_k \widetilde{\mathbf{B}} \\
        \widetilde{\mathbf{B}}^{\intercal} \widetilde{\mathbf{P}}_k \widetilde{\mathbf{A}} & \widetilde{\mathbf{R}} + \widetilde{\mathbf{B}}^{\intercal} \widetilde{\mathbf{P}}_k \widetilde{\mathbf{B}}
    \end{bmatrix} \tilde{\mathbf{z}}_t \nonumber \\
    &\eqqcolon \tilde{\mathbf{z}}_t^{\intercal} \begin{bmatrix}
        [\widetilde{\mathbf{H}}_k]_{11} & [\widetilde{\mathbf{H}}_k]_{12} \\
        [\widetilde{\mathbf{H}}_k]_{21} & [\widetilde{\mathbf{H}}_k]_{22}
    \end{bmatrix} \tilde{\mathbf{z}}_t = \tilde{\mathbf{z}}_t^{\intercal} \widetilde{\mathbf{H}}_k \tilde{\mathbf{z}}_t,
\end{align}
where $\tilde{\mathbf{z}}_{t} = [\tilde{\mathbf{x}}_{t}^\intercal ~ \tilde{\mathbf{u}}_{t}^\intercal]^\intercal$.
Then, the following policy update (due to Hewer) is guaranteed to converge to the optimal \ac{LQR} policy under controllability assumption \cite{hewer1971iterative}:
\begin{gather}
    \label{eq:policy_update_Gd}
    \begin{aligned}[b]
    \widetilde{\mathbf{K}}_{k+1} = - \left( \widetilde{\mathbf{R}} + \widetilde{\mathbf{B}}^{\intercal} \widetilde{\mathbf{P}}_k \widetilde{\mathbf{B}} \right)^{-1} \widetilde{\mathbf{B}}^{\intercal} \widetilde{\mathbf{P}}_k \widetilde{\mathbf{A}} = - [\widetilde{\mathbf{H}}_k]_{22}^{-1} [\widetilde{\mathbf{H}}_k]_{21},
    \end{aligned}
\end{gather}
which is also reconstructed by information in $\widetilde{\mathbf{H}}_k$.
Furthermore, the cost matrix in \eqref{eq:tilded_lyapunov} can also be reconstructed by the same information:
\begin{gather}
    \label{eq:lyap_data_Gd}
    \begin{medsize}
    \begin{aligned}[b]
    \widetilde{\mathbf{P}}_k = [\widetilde{\mathbf{H}}_k]_{11} + [\widetilde{\mathbf{H}}_k]_{12} \widetilde{\mathbf{K}}_k + \widetilde{\mathbf{K}}_k^\intercal [\widetilde{\mathbf{H}}_k]_{21} + \widetilde{\mathbf{K}}_k^\intercal [\widetilde{\mathbf{H}}_k]_{22} \widetilde{\mathbf{K}}_k.
    \end{aligned}
    \end{medsize}
\end{gather}

Hence, $\widetilde{\mathbf{H}}_k$ provides the required information to perform both policy update and policy evaluation steps in a policy iteration algorithm.
We will see that because of the particular structure of our setup, $\widetilde{\mathbf{H}}_k$ enjoys a special block pattern captured by the proposed Patterned monoid, justifying the recovery of the block matrices $X_1$, $X_2$, $Y_1$, $Y_2$, $Z_1$, and $Z_2$ from $\widetilde{\mathbf{H}}_k$ in \ac{D2SPI}.
\ac{D2SPI} leverages this idea to implicitly learn $\widetilde{\mathbf{H}}_k$ from data ({by adapting the idea of \cite{bradtke1994adaptive}}) and exploit these matrix blocks in order to find a suboptimal solution to the main distributed problem in \eqref{eq:distributed_optimization}.

Here, in addition to the policy update from data as in \eqref{eq:policy_update_Gd}, we show that the same information can be used to also learn a gain margin directly from data (see \Cref{prop:gain_margin}). This gain margin is then used to guarantee the stability of the entire network (see \Cref{thm:stabilizability_learning_phase}).

Finally, for technical reasons, recall that the infinite-horizon state-feedback LQR problem with parameters $(A,B,Q,R)$ can be cast as the minimization of 
\begin{equation}\label{eq:lqr-f}
    f_{\Sigma}(K) \coloneqq \mathrm{Tr} \left[ P_K \Sigma \right]
\end{equation}
over the static stabilizing policy $K$, for some initial state distribution with covariance $\Sigma \succ 0$, where $P_K$ is cost matrix associated with $K$ satisfying the following Lyapunov equation \cite{levine1971optimal, maartensson2009gradient, bu2019lqr}:
\[P_K = (A+BK)^\intercal P_K (A+BK) + Q + K^\intercal R K.\]
Herein, we set $\Sigma = \mathrm{I}$ and consider $f_{\mathrm{I}}(K)$.

\subsection{Technical Observations and Main Proofs}
\label{sec:proofs}
 In the remainder of this section, we first restate some well-known technical facts to make the paper self-contained, and then propose a few additional algebraic facts for our analysis whose proofs are deferred to Appendix \ref{app:proof_lemma}. We then continue with the proof of the main results.
 
\begin{lemma}
    \label{lem:algebraic_fact}
    The following relations hold:
    \begin{enumerate}[labelindent=0pt]
        \item \label{item:algebraic_fact.1} (\cite{horn2012matrix}) When $X\succ 0$,
        \begin{align*}
            M^\intercal X N + N^\intercal X M &\succeq -(a M^\intercal X M + \frac{1}{a} N^\intercal X N), \\
            M^\intercal X N + N^\intercal X M &\preceq a M^\intercal X M + \frac{1}{a} N^\intercal X N,
        \end{align*}
        where $M,N\in\mathbb{R}^{n\times m}$ with $n\geq m$ and $a>0$.
        
        \item \label{item:algebraic_fact.2} (\cite[Lyapunov Equation]{lancaster1970explicit}) Suppose that $A\in\mathbb{R}^{n\times n}$ has spectral radius less than 1, i.e., $\rho(A)<1$.
        Then
        \(
            A^\intercal X A + Q - X = 0
        \)
        has a unique solution,
        \(
            X = \sum_{j=0}^\infty (A^\intercal)^j Q A^j.
        \)
        In this case, if $Q \succ 0$, then $X\succ0$.
        
        \item \label{item:algebraic_fact.3} (\cite[Block matrix inverse forluma (0.8.5.6)]{horn2012matrix}) The following identity holds for matrices ${A}$, ${B}$, ${C}$, and ${D}$ with compatible dimensions,
        \begin{gather*}
        \begin{aligned}
            \begin{bmatrix}
                {A} & {B} \\
                {C} & {D}
            \end{bmatrix}^{-1} = \begin{bmatrix}
                {H}^{-1} & -{H}^{-1} {B} {D}^{-1} \\
                - {D}^{-1} {C} {H}^{-1} & {D}^{-1} + {D}^{-1} {C} {H}^{-1} {B} {D}^{-1}
            \end{bmatrix},
        \end{aligned}
        \end{gather*}
        where ${D}$ and ${H} = {A} - {B} {D}^{-1} {C}$ are invertible.
        
        \item \label{item:algebraic_fact.4} (Matrix Inversion Lemma  \cite{woodbury1950inverting}) The following identity holds,
        \begin{gather*}
            (A + U C V)^{-1} = A^{-1} - A^{-1} U (C^{-1} + V A^{-1} U)^{-1} V A^{-1},
        \end{gather*}
        for matrices $A$, $U$, $C$, and $V$ with compatible dimensions where $A$, $C$, and $A + U C V$ are invertible.
    \end{enumerate}
\end{lemma}

Finally, we provide the main technical lemma that streamlines the properties of the Patterned monoid under algebraic manipulation which will be frequently used in the proofs of \Cref{prop:H_structure} and \Cref{thm:convergence}.
\begin{lemma}
    \label{lem:tools_for_analysis}
    Suppose $\mathbf{N}_r \in \mathrm{PM}(r\times n, \mathbb{R})$ for some $n$ and $ r \geq 2$ such that $\mathbf{N}_r = \mathrm{I}_r \otimes \big( A - B \big) + \mathbbm{1}_r \mathbbm{1}_r^\intercal \otimes B, \text{ for some } A \in \mathrm{G L}(n,\mathbb{R}) \cap \mathbb{S}^{n\times n} \text{ and } B \in \mathbb{S}^{n\times n}$.
    Then the following hold:
    \begin{enumerate}[labelindent=0pt]
        \item \label{item:tools_for_analysis.1} $\det(\mathbf{N}_r) = \det(A-B)^{r-1} \det( A + (r-1) B )$.
    
        \item \label{item:tools_for_analysis.2}  If $\mathbf{N}_r \succ 0$, then we have $A+(\ell-1) B \succ 0$ for all $\ell =  0, 1, \cdots, r$. Furthermore, $A-\ell B\left(A+(\ell-1)B\right)^{-1}B$ is invertible for $\ell=1,2,\cdots,r-1$.
    
        \item \label{lem:tools_for_analysis.3} If $\mathbf{N}_r \succ 0$, then $\mathbf{N}_r^{-1} \in \mathrm{PM}(r\times n, \mathbb{R})$, i.e.,
        \begin{align*}
            \mathbf{N}_r^{-1} = \mathrm{I}_r \otimes \big( F_r + G_r \big) - \mathbbm{1}_r \mathbbm{1}_r^\intercal \otimes G_r,
        \end{align*}
        with $F_r$ and $G_r$ defined as,
        \begin{gather*}
        \begin{aligned}
            F_r &= \big( A - (r-1) B \big( A + (r-2) B \big)^{-1} B \big)^{-1}, \\
            G_r &= \big( A + (r-1) B \big)^{-1} B \big( A - B \big)^{-1}.
        \end{aligned}
        \end{gather*}
        
        \item \label{lem:tools_for_analysis.4}
        If $\mathbf{M}_r = \mathrm{I}_r \otimes \big( C - D \big) + \mathbbm{1}_r \mathbbm{1}_r^\intercal \otimes D$ then,
        \begin{multline*}
            \mathbf{N}_r \mathbf{M}_r = \mathrm{I}_r \otimes \big( A - B \big) \big( C - D \big)\\
            + \mathbbm{1}_r \mathbbm{1}_r^\intercal \otimes \Big( B \big( C - D \big) + \big( A - B \big) D + r B D \Big).\\
        \end{multline*}
    \end{enumerate}
\end{lemma}

\subsubsection{Proof of \Cref{prop:lineargroup}}
\label{subsec:lineargroup}

For any Schur stable matrix $A$, and any symmetric positive definite matrix $Q$ there exists a unique positive definite solution $P$ to the discrete Lyapunov equation described by $P = \sum_{j=0}^\infty (A^\intercal)^j Q A^j$ (\Cref{lem:algebraic_fact}.\ref{item:algebraic_fact.2}).
Note, that $\mathrm{PM}(r\times n,\mathbb{R}) \subset \mathbb{S}^{rn\times rn} \cap \mathrm{L}(r\times n, \mathbb{R})$ by construction.
Therefore, since $\mathrm{PM}(r\times n,\mathbb{R})$ is a monoid and $\mathrm{L}(r\times n, \mathbb{R})$ closed under matrix multiplication by \Cref{lem:tools_for_analysis}, each summand falls in $\mathrm{L}(r\times n, \mathbb{R})$ whenever $Q \in \mathrm{PM}(r\times n,\mathbb{R})$.
Also, as the infinite sum preserves the structure, $P \in \mathrm{L}(r\times n, \mathbb{R})$. To show that $P \in \mathrm{PM}(r\times n,\mathbb{R})$ it now suffices to note that $P \succ 0$ whenever $Q \succ 0$ because then every principle submatrix of $P$ has to be positive definite and thus invertible.
Conversely, if $P \in \mathrm{PM}(r\times n,\mathbb{R})$, then $Q = P - A^\intercal P A^\intercal$ also must lie in $\mathrm{PM}(r\times n,\mathbb{R})$ as $Q \succ 0$.
This completes the proof$.\hfill \square$\\


\subsubsection{Proof of \Cref{prop:H_structure}}
\label{subsec:H_structure}
At iteration $k$ of the learning phase in \Cref{alg:distributed_control_algorithm},
the first claim is a direct consequence of the structure of $\mathcal{G}_{d,\mathrm{learn}}$ during the learning phase, where $\mathcal{G}_{d,\mathrm{learn}} = \mathcal{K} \left( \mathcal{G}_d \right)$ and hence $\u_i = (K_k - L_k) \x_i + L_k \sum_{j \in \mathcal{V}_{\mathcal{G}_{d,\mathrm{learn}}}} \x_j$ for all $i\in\mathcal{V}_{\mathcal{G}_{d,\mathrm{learn}}}$, which, in turn, results in $\tilde{\mathbf{u}}_t = \widetilde{\mathbf{K}}_k \tilde{\mathbf{x}}_t$ with $\widetilde{\mathbf{K}}_k$ as claimed. The stability of the policy $\widetilde{\mathbf{K}}_k$ for the pair $(\widetilde{\mathbf{A}},\widetilde{\mathbf{B}})$ throughout the learning phase is then argued in \Cref{subsec:convergence} under \Cref{assmp:main}.

Next, the cost matrix $\widetilde{\mathbf{P}}_k$ associated with $\widetilde{\mathbf{K}}_k$ satisfies the Lyapunov equation \eqref{eq:tilded_lyapunov}.
We can verify that $\widetilde{\mathbf{A}} + \widetilde{\mathbf{B}} \widetilde{\mathbf{K}}_k \in \mathrm{L}(d\times n, \mathbb{R})$ which is also Schur stable, and $\widetilde{\mathbf{Q}} + \widetilde{\mathbf{K}}_k^\intercal \widetilde{\mathbf{R}} \widetilde{\mathbf{K}}_k \in \mathrm{PM}(d\times n, \mathbb{R})$. Thus, by \Cref{prop:lineargroup}, we conclude that $\widetilde{\mathbf{P}}_k \in \mathrm{PM}(d\times n, \mathbb{R})$. 
So, let
\begin{gather}
    \label{eq:P_tilde_structure}
    \widetilde{\mathbf{P}}_k = \mathrm{I}_d \otimes \big( P_1 - P_2 \big) + \mathbbm{1} \mathbbm{1}^{\intercal} \otimes P_2,
\end{gather}
for some $P_1, P_2$. Note that  $\widetilde{\mathbf{K}}_k$ is stabilizing and $\widetilde{\mathbf{Q}} + \widetilde{\mathbf{K}}_k^\intercal \widetilde{\mathbf{R}} \widetilde{\mathbf{K}}_k \succ 0$, therefore $\widetilde{\mathbf{P}}_k \succ 0$ from \eqref{eq:tilded_lyapunov}. But then, by \Cref{lem:tools_for_analysis}.\Cref{item:tools_for_analysis.2} and the structure of $\widetilde{\mathbf{P}}_k$ from \eqref{eq:P_tilde_structure}, we claim that $\Delta P_k := P_1 - P_2 \succ 0$. Next, one can also verify that
\begin{gather*}
\begin{aligned}
    \big( \widetilde{\mathbf{A}} + &\widetilde{\mathbf{B}} \widetilde{\mathbf{K}}_k \big)^{\intercal} \widetilde{\mathbf{P}}_k \big( \widetilde{\mathbf{A}} + \widetilde{\mathbf{B}} \widetilde{\mathbf{K}}_k \big)\\
    &= \mathrm{I}_d \otimes \big( A_{\Delta K_k}^\intercal (\Delta P_k) A_{\Delta K_k} \big) + \mathbbm{1} \mathbbm{1}^{\intercal} \otimes (\star),\\
    \widetilde{\mathbf{Q}} + &\widetilde{\mathbf{K}}_k^\intercal \widetilde{\mathbf{R}} \widetilde{\mathbf{K}}_k \\
    &=
    \mathrm{I}_d \otimes \big( Q_d + (\Delta K_k)^\intercal R (\Delta K_k) \big) + \mathbbm{1} \mathbbm{1}^{\intercal} \otimes (\star),
\end{aligned}
\end{gather*}
where $\Delta K_k = K_k-L_k$, $A_{\Delta K_k} = A + B (\Delta K_k)$, $Q_d = Q_1 + d Q_2$, and $(\star)$ is hiding extra terms.
But then, by \eqref{eq:tilded_lyapunov} and \eqref{eq:P_tilde_structure} we obtain that $\Delta P_k$ must satisfy:
\begin{gather}
    \label{eq:new_lyap}
    \Delta P_k = A_{\Delta K_k}^\intercal (\Delta P_k) A_{\Delta K_k} + Q_d + \Delta K_k^\intercal R \Delta K_k,
\end{gather}
which itself is a Lyapunov equation. Finally, since $Q_d + \Delta K_k^\intercal R \Delta K_k \succ 0$ and $\Delta P_k \succ 0$, by Lyapunov Stability Criterion we conclude that $A_{\Delta K_k}$ is Schur stable. This completes the proof.
$\hfill\square$\\

\subsubsection{Proof of \Cref{prop:gain_margin}}
\label{subsec:gain_margin}
Define the Lyapunov candidate function $V_k(\x_t) = \x_t^\intercal \Delta P_k \x_t$ with $\Delta P_k \succ 0$ as in \eqref{eq:new_lyap} and where $\x_t$ contains the states of the closed-loop system $\x_{t+1} = \big( A + B (K_k - \alpha L_k) \big) \x_t$ for some scalar $\alpha$.
We show that for the given choice of $\alpha$, $V_k$ is decreasing.
We define
\begin{gather*}
    \Delta V_k(\x_t) \coloneqq V_k(\x_{t+1}) - V_k(\x_t) = \x_t^\intercal \Gamma_k \x_t,
\end{gather*}
where
\begin{gather*}
    \Gamma_k = \left( A_{\Delta K_k} + (1 - \alpha) L_k \right)^\intercal \Delta P_k \left( A_{\Delta K_k} + (1 - \alpha) L_k \right) - \Delta P_k.
\end{gather*}
Suppose $\alpha < 1$, then from \eqref{eq:new_lyap}, 
\begin{gather*}
\begin{aligned}
    \Gamma_k 
    =&- (Q_d + \Delta K_k^\intercal R \Delta K_k) + (1-\alpha)^2 L_k^\intercal B^\intercal \Delta P_k B L_k  \\
    &+ (1-\alpha) \left( A_{\Delta K_k}^\intercal \Delta P_k B L_k + L_k^\intercal B^\intercal \Delta  P_k A_{\Delta K_k} \right) \\
    \preceq &- (Q_d + \Delta K_k^\intercal R \Delta K_k) + (1-\alpha)^2 L_k^\intercal B^\intercal \Delta P_k B L_k \\ &+(1-\alpha) \left( a A_{\Delta K_k}^\intercal \Delta P_k A_{\Delta K_k} + (1/a) L_k^\intercal B^\intercal \Delta  P_k B L_k \right) \\
    =& - \big( Q_d + \Delta K_k^\intercal R \Delta K_k \big) +(1-\alpha) a A_{\Delta K_k}^\intercal \Delta P_k A_{\Delta K_k}  \\
    &+(1-\alpha) \left( 1/a + 1 - \alpha \right) L_k^\intercal B^\intercal \Delta P_k B L_k
\end{aligned}
\end{gather*}
where the inequality holds for any $a>0$ due to \Cref{lem:algebraic_fact}.
Let $\beta = (1-\alpha)/2$ and choose $a = \beta + \sqrt{\beta^2 + 1}$.
Then, $\left( 1/a + 1 - \alpha \right) = a$ and thus
\begin{multline*}
    \lambda_{\max}(\Gamma_k) 
    \leq - \lambda_{\min} \big[ Q_d + \Delta K_k^\intercal R \Delta K_k \big] +\\
    (1-\alpha) a \lambda_{\max} \big[ A_{\Delta K_k}^\intercal \Delta P_k A_{\Delta K_k} + L_k^\intercal B^\intercal \Delta P_k B L_k \big]
\end{multline*}
Now, using the parameters \eqref{eq:XYZ} constructing the blocks of \eqref{eq:H_has_blocks}, we obtain
\begin{gather*}
\begin{aligned}
        L_k^\intercal B^\intercal \Delta P_k B L_k = & L_k^\intercal (\Delta Y - R) L_k,\\
        A_{\Delta K_k}^\intercal \Delta P_k A_{\Delta K_k} = & \Delta X - Q_d + \Delta K_k^\intercal \Delta Z + \Delta Z^\intercal \Delta K_k\\
        &+ \Delta K_k^\intercal \big( \Delta Y - R \big) \Delta K_k\eqqcolon \Xi_k.
\end{aligned}
\end{gather*}
Thus, the latter bound can be obtained completely from data as
\begin{gather*}
     \lambda_{\max}(\Gamma_k) 
    \leq \left( (1-\alpha) a - \gamma_k \right) \lambda_{\max} \left[ \Xi_k + L_k^\intercal (\Delta Y - R) L_k \right]
\end{gather*}
with
\begin{gather*}
    \gamma_k \coloneqq \lambda_{\min} \big[ Q_d + \Delta K_k^\intercal R \Delta K_k \big]\big/\lambda_{\max} \left[ \Xi_k + L_k^\intercal (\Delta Y - R) L_k \right],
\end{gather*}
which coincides with updates in
\Cref{step:singular_values} of \Cref{alg:distributed_control_algorithm}.

Finally, from the hypothesis $1-\tau_k < \alpha < 1$ with $\tau_k = \sqrt{\gamma_k^2 / (1 + \gamma_k)}$, we obtain that $\gamma_k^2 - 4\beta^2 \gamma_k - 4\beta^2 > 0$.
But, since $\gamma_k > 0$, this second-order term in $\gamma_k$ is positive only if
\begin{gather*}
    \gamma_k > 2\beta^2 + 2\beta \sqrt{\beta^2 + 1} = 2\beta (\beta + \sqrt{\beta^2 + 1}) = (1-\alpha) a.
\end{gather*}
Therefore, $\Delta V_k(\x_t) < 0$ for $1-\tau_k < \alpha < 1$.
Similar reasoning for $1 < \alpha < 1+\tau_k$ also shows that $\Delta V_k(\x_t) < 0$ which completes the proof. $\hfill\square$\\


\subsubsection{Proof of \Cref{thm:stabilizability_learning_phase}}
\label{subsec:stabilizability_learning_phase}
From \Cref{def:policy} (according to a consistent choice of labeling of the nodes so that the last $d$ nodes are chosen as $\mathcal{G}_{d}$) and \Cref{prop:H_structure}, the feedback policy in  \cref{line:policy_update} of \Cref{alg:distributed_control_algorithm} can be cast in the compact form,
\begin{gather*}
\begin{aligned}
    \mathrm{Policy}_k (\mathcal{V}_\mathcal{G}) &= \begin{bmatrix}
        \mathrm{Policy}_k ( \mathcal{V}_{\mathcal{G} \setminus \mathcal{G}_{d,\mathrm{learn}}} ) \\ \mathrm{Policy}_k( \mathcal{V}_{\mathcal{G}_{d,\mathrm{learn}}} )
    \end{bmatrix} \\
    &= \left[ \begin{array}{cc}
     \hat{\mathbf{K}}_{\mathcal{G} \setminus \mathcal{G}_{d}} & \hat{\mathbf{L}}_{\mathcal{G} \setminus \mathcal{G}_{d}}
     \\ \rule{0pt}{1.2\normalbaselineskip}
    \mathbf{0} & \widetilde{\mathbf{K}}_k
\end{array} \right] \begin{bmatrix}
        \mathrm{State}_k ( \mathcal{V}_{\mathcal{G} \setminus \mathcal{G}_{d,\mathrm{learn}}} ) \\ \mathrm{State}_k( \mathcal{V}_{\mathcal{G}_{d,\mathrm{learn}}} )
    \end{bmatrix} \\
    &\eqqcolon \hat{\mathbf{K}}_k \mathrm{State}_k(\mathcal{G}),
\end{aligned}
\end{gather*}
where $\hat{\mathbf{L}}_{\mathcal{G} \setminus \mathcal{G}_{d}} \coloneqq   \frac{\tau_k}{d-1} [\mathcal{A}_{\mathcal{G}}]_{12} \otimes L_k, \;
    \widetilde{\mathbf{K}}_k \coloneqq \mathrm{I}_d \otimes ( K_k - L_k ) + \mathbbm{1}_d \mathbbm{1}_d^\intercal \otimes L_k, \; \hat{\mathbf{K}}_{\mathcal{G} \setminus \mathcal{G}_{d}} \coloneqq \mathrm{I}_{N-d} \otimes K_k - \left( \mathrm{I}_{N-d} - \frac{\tau_k}{d-1} \mathcal{A}_{\mathcal{G} \setminus \mathcal{G}_d} \right) \otimes L_k,$
$\mathcal{A}_{\mathcal{G}}$ denotes the adjacency matrix of $\mathcal{G}$, and $[\mathcal{A}_{\mathcal{G}}]_{12}$ is its submatrix capturing the interconnection of $\mathcal{G} \setminus \mathcal{G}_{d}$ and $\mathcal{G}_{d}$:
\begin{gather*}
    \resizebox{0.45\columnwidth}{!}{
    $\begin{aligned}
        \mathcal{A}_{\mathcal{G}} = \begin{bmatrix}
    \mathcal{A}_{\mathcal{G} \setminus \mathcal{G}_{d}} & [\mathcal{A}_{\mathcal{G}}]_{12}\\
    * & \mathcal{A}_{\mathcal{G}_{d}}
\end{bmatrix}.
    \end{aligned}$
    }
\end{gather*}
Note that the structure of $\hat{\mathbf{K}}_k$ emanates from the fact that the information exchange is unidirectional during the learning phase.
Now consider the closed-loop system of $\mathcal{G}$,
\begin{gather*}
\resizebox{0.99\columnwidth}{!}{
  $  \begin{aligned}
        &\hat{\mathbf{A}}_{\mathcal{G}} |_\textit{cl} \coloneqq \hat{\mathbf{A}} + \hat{\mathbf{B}} \hat{\mathbf{K}}_k \hspace{6cm}\\
    &=
    \left[ \begin{array}{cc}
         \mathrm{I}_{N-d} \otimes A + \big( \mathrm{I}_{N-d} \otimes B \big) \hat{\mathbf{K}}_{\mathcal{G} \setminus \mathcal{G}_d} & \big( \mathrm{I}_{N-d} \otimes B \big) \hat{\mathbf{L}}_{\mathcal{G} \setminus \mathcal{G}_d}\\
         \mathbf{0} & \widetilde{\mathbf{A}}_{\widetilde{\mathbf{K}}_k}
    \end{array} \right],
    \end{aligned}$
    }
\end{gather*}
where $ \widetilde{\mathbf{A}}_{\widetilde{\mathbf{K}}_k} \coloneqq \widetilde{\mathbf{A}} + \widetilde{\mathbf{B}} \widetilde{\mathbf{K}}_k$ is the closed-loop system of $\mathcal{G}_d$.
Define $S = \mathrm{I}_{N-d} - \frac{\tau_k}{d-1} \mathcal{A}_{\mathcal{G} \setminus \mathcal{G}_d}$ and let $J$ be the Jordan form of $S$ according to the similarity transformation $\mathcal{T} \in\mathbb{R}^{(N-d)\times(N-d)}$ such that $\mathcal{T} S \mathcal{T}^{-1} = J$.
Now consider the following similarity transformation of $\hat{\mathbf{A}}_{\mathcal{G}} |_\textit{cl}$,
\begin{equation*}
\resizebox{\columnwidth}{!}{
$\begin{aligned}
    &\begin{bmatrix}
        \mathcal{T} \otimes \mathrm{I}_{n} & \mathbf{0} \\
        \mathbf{0} & \mathrm{I}_{d} \otimes \mathrm{I}_{n}
    \end{bmatrix}
    \Big( \hat{\mathbf{A}}_{\mathcal{G}} |_\textit{cl} \Big)
    \begin{bmatrix}
        \mathcal{T} \otimes \mathrm{I}_{n} & \mathbf{0} \\
        \mathbf{0} & \mathrm{I}_{d} \otimes \mathrm{I}_{n}
    \end{bmatrix}^{-1} \\
    &= \left[ \begin{array}{cc}
        \mathrm{I}_{N-d} \otimes \big( A + B K_k \big) - J \otimes B L_k & \mathbf{*}\\
        \mathbf{0} & \widetilde{\mathbf{A}}_{\widetilde{\mathbf{K}}_k} 
    \end{array} \right].
\end{aligned}$
}
\end{equation*}
Note that $\mathrm{I}_{N-d} \otimes \big( A + B K_k \big) - J \otimes B L_k$ is a block upper triangular matrix whose diagonal blocks are equal to $A + B ( K_k - \lambda_i(S) L_k )$ for $i=1,\cdots,N-d$.
We already know from \Cref{prop:H_structure} that $ \widetilde{\mathbf{A}}_{\widetilde{\mathbf{K}}_k}$ is Schur stable.
Hence, in order to show that $\hat{\mathbf{A}}_{\mathcal{G}} |_\textit{cl}$ is Schur stable, it suffices to show that $\rho \left( A + B ( K - \lambda_i(S) L ) \right) < 1$ for $i=1,\cdots,N-d$.
Recall that $\left\vert \lambda_i(\mathcal{A}_\mathcal{G}) \right\vert \leq d_{\max}$ \cite{bollobas2013modern}, thus by definition of $S$ and the fact that $d_{\max}=d-1$, we conclude that $\left\vert \lambda_i(S) - 1 \right\vert \leq \tau_k$.
The rest of the proof now follows directly from \Cref{prop:gain_margin}. $\hfill\square$\\

\subsubsection{Proof of \Cref{thm:convergence}}
\label{subsec:convergence}
At iteration $k$ of the learning phase in \Cref{alg:distributed_control_algorithm}, consider $\widetilde{\mathbf{H}}_k$ and its corresponding blocks as defined in \eqref{eq:H_has_blocks}.
First, we consider the structure of the stabilizing feedback policy $\widetilde{\mathbf{K}}_k$ as shown in \Cref{prop:H_structure}, together with that of system parameters $(\widetilde{\mathbf{A}},\widetilde{\mathbf{B}},\widetilde{\mathbf{Q}}, \widetilde{\mathbf{R}})$, and apply \Cref{lem:tools_for_analysis} and \Cref{prop:lineargroup} to conclude that $[\widetilde{\mathbf{H}}_k]_{11},[\widetilde{\mathbf{H}}_k]_{22} \in \mathrm{PM}(d\times n, \mathbb{R})$ and $[\widetilde{\mathbf{H}}_k]_{21} \in \mathrm{L}(d\times n, \mathbb{R})$. Thus, we get
\begin{gather*}
\begin{aligned}
        [\widetilde{\mathbf{H}}_k]_{11} &= \mathrm{I}_d \otimes \big( X_1 - X_2 \big) + \mathbbm{1} \mathbbm{1}^{\intercal} \otimes X_2,\\
        [\widetilde{\mathbf{H}}_k]_{22} &= \mathrm{I}_d \otimes \big( Y_1 - Y_2 \big) + \mathbbm{1} \mathbbm{1}^{\intercal} \otimes Y_2,\\
        [\widetilde{\mathbf{H}}_k]_{21} &=  \mathrm{I}_d \otimes \big( Z_1 - Z_2 \big) + \mathbbm{1} \mathbbm{1}^{\intercal} \otimes Z_2,
    \end{aligned}
\end{gather*}
which coincides with the recovery of $X_i,Y_i$ and $Z_i$ for $i=1,2$ in \Cref{step:blocks} of \Cref{alg:distributed_control_algorithm}.
We can also unravel the structure of block matrices constructing $\widetilde{\mathbf{H}}_k$ and obtain that
\begin{gather}\label{eq:XYZ}
\resizebox{\columnwidth}{!}{
$\begin{aligned}
    X_1 &= Q_1 + (d-1)Q_2 + A^\intercal P_1 A, &\;
    Y_2 &= B^\intercal P_2 B,\\
    X_2 &= -Q_2 + A^\intercal P_2 A, &\;
    Z_1 &= B^\intercal P_1 A,\\
    Y_1 &= R + B^\intercal P_1 B, &\;
    Z_2 &= B^\intercal P_2 A,
\end{aligned}$
}
\end{gather}
with $P_1$ and $P_2$ as in \eqref{eq:P_tilde_structure}.
Then, by \Cref{lem:tools_for_analysis}, we similarly get:
\begin{align*}
 [\widetilde{\mathbf{H}}_k]_{22}^{-1} = \mathrm{I}_d \otimes (F - G) + \mathbbm{1} \mathbbm{1}^{\intercal} \otimes G,
\end{align*}
where (by \Cref{lem:tools_for_analysis} \Cref{lem:tools_for_analysis.3}) $F$ and $G$ must satisfy 
\begin{align*}
    F^{-1} &=  Y_1 - (d-1) Y_2 \big( Y_1 + (d-2) Y_2 \big)^{-1} Y_2 , \\
    G &= \Big( Y_1 + (d-1)Y_2 \Big)^{-1} Y_2 \Big( Y_1 - Y_2 \Big)^{-1},
\end{align*}
which coincides with the definitions in \Cref{step:inverses} of \Cref{alg:distributed_control_algorithm}. Finally, by definition of $K_{k+1}$ and $L_{k+1}$ in \Cref{step:inverses} and \Cref{lem:tools_for_analysis} \Cref{lem:tools_for_analysis.4}, one can verify that
$\widetilde{\mathbf{K}}_{k+1} = - [\widetilde{\mathbf{H}}_k]_{22}^{-1} [\widetilde{\mathbf{H}}_k]_{21}$,
which coincides with the policy iteration in the Hewer's algorithm \cite{hewer1971iterative} for the system in $\mathcal{G}_{d,\text{learn}}$ (see also \Cref{subsec:analysis_setup}). Note that by assumption the pair $(A,B)$ is controllable, so is the system $(\widetilde{\mathbf{A}},\widetilde{\mathbf{B}})$ in $\mathcal{G}_{d,\text{learn}}$.
Therefore, these updates are guaranteed to remain stabilizing and converge to the claimed optimal \ac{LQR} policy $\widetilde{\mathbf{K}}^*$ provided that we have access to the true parameters $\widetilde{\mathbf{H}}_k$.

Next, consider \ac{LQR} cost $f_{\mathrm{I}}(\widetilde{\mathbf{K}})$ as in \eqref{eq:lqr-f} but for the problem parameters $(\widetilde{\mathbf{A}},\widetilde{\mathbf{B}},\widetilde{\mathbf{Q}}, \widetilde{\mathbf{R}})$.
For completing the proof of convergence, it is left to argue that there exists a large enough integer $C$ such that, at each iteration $k$ of the learning phase, the recursive least square in \Cref{alg:SPE} provides a more accurate estimation, denoted by $\otheta_k^\circ$, of the true parameters $\otheta_k = \mathrm{vech}(\widetilde{\mathbf{H}}_k)$, and the \ac{LQR} cost $f_{\mathrm{I}}(\widetilde{\mathbf{K}}) = \mathrm{Tr}(\widetilde{\mathbf{P}}_{\widetilde{\mathbf{K}}})$ decreases. This claim essentially follows by \cite[Theorem 1]{bradtke1994adaptive} which we try to summarize for completeness.
For that, at iteration $k$, let us denote the policy obtained using the estimated parameters by $\widetilde{\mathbf{K}}_{k}^\circ$ which in turn estimates the true policy $\widetilde{\mathbf{K}}_{k}$.
We then define a ``Lyapunov'' function candidate
\[s_k \coloneqq f_{\mathrm{I}}(\widetilde{\mathbf{K}}_{k-1}^\circ) + \big\| \otheta_{k-2} - \otheta_{k-2}^\circ \big\|.\]
Following the same induction reasoning as that of \cite[Theorem 1]{bradtke1994adaptive} and under persistently exciting input, there exists an integer $C$ such that
\begin{gather*}
    s_{k+1} = s_{k} - \varepsilon_1(C) \big\| \otheta_{k-1} - \otheta_{k-1}^\circ \big\| - \varepsilon_2(C) \big\| \widetilde{\mathbf{K}}_{k} - \widetilde{\mathbf{K}}_{k-1}^\circ \big\|^2,
\end{gather*}
for some positive constants $\varepsilon_1(C)$ and $\varepsilon_1(C)$ that are independent of $k$. But then, $s_{k+1} \leq s_{k}$ and
\begin{gather*}
    \varepsilon_1(C) \sum_{k=2}^\infty \big\| \otheta_{k-1} - \otheta_{k-1}^\circ \big\| \leq s_1;\;\; \varepsilon_2(C) \sum_{k=2}^\infty \big\| \widetilde{\mathbf{K}}_{k} - \widetilde{\mathbf{K}}_{k-1}^\circ \big\|^2 \leq s_1.
\end{gather*}
Also, $s_1$ is bounded as $\widetilde{\mathbf{K}}_{1}$ stabilizes $(\widetilde{\mathbf{A}}, \widetilde{\mathbf{B}})$. This guarantees that, first, $\widetilde{\mathbf{K}}_{k}^\circ$ remains stabilizing as $f_{\mathrm{I}}(\widetilde{\mathbf{K}}_{k-1}^\circ) \leq s_0$ and $\widetilde{\mathbf{Q}} \succ 0$; second, the estimates $\otheta_{k-1}^\circ$ become more accurate; and third, $\widetilde{\mathbf{K}}_{k}^\circ \to \widetilde{\mathbf{K}}^*$ as $\widetilde{\mathbf{K}}_{k+1} \to \widetilde{\mathbf{K}}^*$ at $k \to \infty$ \cite{hewer1971iterative}. This completes the proof. $\hfill\square$\\


\subsubsection{Proof of \Cref{thm:suboptimality}}
\label{subsec:suboptimality}
Consider the \ac{LQR} cost $f_{\mathrm{I}}(\hat{\mathbf{K}})$ as in \eqref{eq:lqr-f} but for the problem parameters $(\hat{\mathbf{A}},\hat{\mathbf{B}},\hat{\mathbf{Q}}, \hat{\mathbf{R}})$.
In the ``unstructured'' case (i.e. ignoring the constraint $K\in \mathcal{U}_{m,n}^N(\mathcal{G})$), we know that the optimal \ac{LQR} cost matrix for the entire networked system satisfies \cite{goodwin2001control},
\begin{align}
    \label{eq:lyap_genearl_LQR}
    \hat{\mathbf{P}}_{\mathrm{lqr}}^* = \hat{\mathbf{A}}_{\hat{\mathbf{K}}_{\mathrm{lqr}}}^\intercal \hat{\mathbf{P}}_{\mathrm{lqr}} \hat{\mathbf{A}}_{\hat{\mathbf{K}}_{\mathrm{lqr}}} + \hat{\mathbf{K}}_{\mathrm{lqr}}^{\intercal} \hat{\mathbf{R}} \hat{\mathbf{K}}_{\mathrm{lqr}} + \hat{\mathbf{Q}},
\end{align}
where $\hat{\mathbf{K}}_{\mathrm{lqr}} = \argmin_{K} f_{\mathrm{I}}(K)$.
Moreover, the cost matrix $\hat{\mathbf{P}}^*$-- associated with the structured policy $\hat{\mathbf{K}}^*$ learned by \Cref{alg:distributed_control_algorithm}--satisfies
\begin{align}
    \label{eq:lyap_our_solution}
    \hat{\mathbf{P}}^* = \mathbf{A}_{\hat{\mathbf{K}}^*}^\intercal \hat{\mathbf{P}}^* \mathbf{A}_{\hat{\mathbf{K}}^*} + \hat{\mathbf{K}}^{*\intercal} \hat{\mathbf{R}} \hat{\mathbf{K}}^* + \hat{\mathbf{Q}},
\end{align}
where $\mathbf{A}_{\hat{\mathbf{K}}^*} = \hat{\mathbf{A}} + \hat{\mathbf{B}} \hat{\mathbf{K}}^*$ and $\hat{\mathbf{K}}^* \in \mathcal{U}_{m,n}^N(\mathcal{G})$.
Finally, let $\hat{\mathbf{K}}_{\mathrm{struc}}^* \in \argmin_{K\in \mathcal{U}_{m,n}^N(\mathcal{G})} f_{\mathrm{I}}(K)$ denote a ``structured'' stabilizing optimal \ac{LQR} policy which is associated with the cost matrix $\hat{\mathbf{P}}_{\mathrm{struc}}^*$. We know such a policy exists since the smooth cost is lower-bounded and $\hat{\mathbf{K}}_1 =  I_N \otimes K_1 \in \mathcal{U}_{m,n}^N(\mathcal{G})$ is a feasible point of this optimization---as $K_1$ is assumed to be stabilizing for the single pair $(A,B)$.
Therefore,
\begin{gather*}
    \mathrm{Tr} \left[ \hat{\mathbf{P}}_{\mathrm{lqr}}^* \right] = f_{\mathrm{I}} \left( \hat{\mathbf{K}}_{\mathrm{lqr}}^* \right) \leq f_{\mathrm{I}} \left( \hat{\mathbf{K}}_{\mathrm{struc}}^* \right) \leq f_{\mathrm{I}} \left( \hat{\mathbf{K}}^* \right) = \mathrm{Tr} \left[ \hat{\mathbf{P}}^* \right],
\end{gather*}
where the last inequality above follows by the fact that $\hat{\mathbf{K}}^*$ is a feasible solution to the structured problem by construction, \textit{i.e.}, $\hat{\mathbf{K}}^* \in \mathcal{U}_{m,n}^N(\mathcal{G})$.
Therefore, $0 \leq \mathrm{gap} (\hat{\mathbf{K}}^*) \leq \mathrm{Tr} \left[ \hat{\mathbf{P}}^* - \hat{\mathbf{P}}_{\mathrm{lqr}}^* \right]$.
But then, one can obtain from \eqref{eq:lyap_genearl_LQR}, \eqref{eq:lyap_our_solution}, and some algebraic manipulation,
\begin{align*}
    \hat{\mathbf{P}}^* - \hat{\mathbf{P}}_{\mathrm{lqr}}^* = \mathbf{A}_{\hat{\mathbf{K}}_{\mathrm{lqr}}^*}^\intercal \Big( \hat{\mathbf{P}}^* - \hat{\mathbf{P}}_{\mathrm{lqr}} \Big) \mathbf{A}_{\hat{\mathbf{K}}_{\mathrm{lqr}}} + \mathbf{M}',
\end{align*}
where
\begin{gather*}
    \mathbf{M}' = \mathbf{A}_{\hat{\mathbf{K}}^{*}}^\intercal \hat{\mathbf{P}}^{*} \mathbf{A}_{\hat{\mathbf{K}}^{*}} - \mathbf{A}_{\hat{\mathbf{K}}_{\mathrm{lqr}}^{*}}^{\intercal} \hat{\mathbf{P}}^{*} \mathbf{A}_{\hat{\mathbf{K}}_{\mathrm{lqr}}^{*}} + \hat{\mathbf{K}}^{*\intercal} \hat{\mathbf{R}} \hat{\mathbf{K}}^{*} - \hat{\mathbf{K}}^{\intercal}_{\mathrm{lqr}} \hat{\mathbf{R}} \hat{\mathbf{K}}_{\mathrm{lqr}}.
\end{gather*}
Since $\hat{\mathbf{P}}^* - \hat{\mathbf{P}}_{\mathrm{lqr}}^* \succ 0$ and $\hat{\mathbf{A}}_{\hat{\mathbf{K}}_{\mathrm{lqr}}}$ is contractible by the hypothesis, from the first part and Theorem 1 in \cite{mori1982discrete} we obtain,
\begin{align*}
     \mathrm{gap} (\hat{\mathbf{K}}^*) \leq \frac{\mathrm{Tr}(\mathbf{M}')}{1-\sigma_{\max}^2 \big( \hat{\mathbf{A}}_{\hat{\mathbf{K}}_{\mathrm{lqr}}} \big) } = \frac{\mathrm{Tr}(\mathbf{M})}{1-\sigma_{\max}^2 \big( \hat{\mathbf{A}}_{\hat{\mathbf{K}}_{\mathrm{lqr}}} \big) },
\end{align*} 
where the last equality follows by the cyclic permutation property of trace and definition of $\mathbf{A}_{\hat{\mathbf{K}}^{*}}$. $\hfill\square$

\section{Proof of \Cref{lem:tools_for_analysis}}
\label{app:proof_lemma}

First, we show that the following algebraic identities hold which will be used in the proof of \Cref{lem:tools_for_analysis}.
\begin{lemma}
    \label{lem:algebraic_identities}
    Suppose $A$ and $B$ are symmetric matrices such that $A$, $A-B$, and $A+(n-1) B$ are all invertible for some integer $n$.
    Then the following relations hold:
    \begin{enumerate}[labelindent=0pt]
        \item \label{lem:algebraic_identities.1} $\big( A + n B \big) \big( A + (n-1) B \big)^{-1} \big( A - B \big) = A - n B \big( A + (n-1) B \big)^{-1} B$.
        
        \item \label{lem:algebraic_identities.2} $\big( A + n B \big) \big( A + (n-1) B \big)^{-1} \big( A - B \big) = \big( A - B \big) \big( A + (n-1) B \big)^{-1} \big( A + n B \big)$.
        
        \item \label{lem:algebraic_identities.3} $\big( A + n B \big) \big( A-B \big)^{-1} B = B \big( A-B \big)^{-1} \big( A + n B \big)$.
    \end{enumerate}
\end{lemma}

\textit{Proof:} These claims follow by the algebraic manipulations below: First,
\begin{gather*}
\begin{aligned}
\big( &A + n B \big) \big( A + (n-1) B \big)^{-1} \big( A - B \big)\\
    =& \Big( \big( A + (n-1) B \big) + B \Big) \big( A + (n-1) B \big)^{-1} \big( A - B \big) \\
    =& \Big( \mathrm{I} + B \big( A + (n-1) B \big)^{-1} \Big) \big( A - B \big) \\
    =& A - B + B \big( A + (n-1) B \big)^{-1} A - B \big( A + (n-1) B \big)^{-1} B \\
    =& A + B \Big( \big( A + (n-2) B \big)^{-1} A - \mathrm{I} \Big) - B \big( A + (n-1) B \big)^{-1} B \\
    =& A + B \big( A + (n-2) B \big)^{-1} \Big( A - \big( A + (n-2) B \big) \Big) \\
    &- B \big( A + (n-1) B \big)^{-1} B \\
    =& A - (n-1) B \big( A + (n-1) B \big)^{-1} B - B \big( A + (n-1) B \big)^{-1} B \\
    =& A - n B \big( A + (n-1) B \big)^{-1} B.
\end{aligned}
\end{gather*}
Second,
\begin{gather*}
\begin{aligned}
    \big( &A + n B \big) \big( A + (n-1) B \big)^{-1} \big( A - B \big)\\
    =& -n \big( A - B \big) \big( A + (n-1) B \big)^{-1} \big( A - B \big) \\
    &+ (n+1) A \big( A + (n-1) B \big)^{-1} \big( A - B \big) \\
    =& -n \big( A - B \big) \big( A + (n-1) B \big)^{-1} \big( A - B \big) \\
    &+ (n+1) \big( \mathrm{I} + (n-1) B A^{-1} \big)^{-1} \big( A - B \big) \\
    =& -n \big( A - B \big) \big( A + (n-1) B \big)^{-1} \big( A - B \big) \\
    &+ (n+1) \big( A - B \big) \Big( \big( \mathrm{I} + (n-1) B A^{-1} \big) \big( A - B \big) \Big)^{-1} \big( A - B \big) \\
    =& -n \big( A - B \big) \big( A + (n-1) B \big)^{-1} \big( A - B \big) \\
    &+ (n+1) \big( A - B \big) \Big( \big( A - B \big) \big( \mathrm{I} + (n-1) A^{-1} B \big) \Big)^{-1} \big( A - B \big) \\
    =& -n \big( A - B \big) \big( A + (n-1) B \big)^{-1} \big( A - B \big) \\
    &+ (n+1) \big( A - B \big) \big( A + (n-1) B \big)^{-1} A \\
    =& \big( A - B \big) \big( A + (n-1) B \big)^{-1} \big( A + n B \big).
\end{aligned}
\end{gather*}
Finally,
\begin{gather*}
    \begin{aligned}
    \big( &A + n B \big) \big( A - B \big)^{-1} B \\
    &= \big( A - B + (n+1) B \big) \big( A-B \big)^{-1} B\\
    &= \big( \mathrm{I} + (n+1) B (A-B)^{-1} \big) B = B \big( \mathrm{I} + (n+1) (A-B)^{-1} B \big) \\
    &= B \big( A - B \big)^{-1} \big( A + n B \big).
    \end{aligned}
\end{gather*}
$\hfill\square$

\subsubsection*{Proof of \Cref{lem:tools_for_analysis}}
\label{subsec:tools_for_analysis}
Part 1) We prove the claim by induction.
First, note that both $\mathbf{N}_r$ and its principle submatrix $A$ are invertible.
For $r=2$, by Schur complement of $\mathbf{N}_2$, we get
\begin{gather*}
    \begin{aligned}
        \det(\mathbf{N}_2) &= \det(A) \det(A - B A^{-1} B) \\
        &= \det(A) \det(\mathrm{I} - B A^{-1}) \det(\mathrm{I} + B A^{-1}) \det(A) \\
        &= \det(A - B) \det(A + B).
    \end{aligned}
\end{gather*}
Now, suppose the claim holds for $r=p$.
Then, for $r=p+1$, similarly by Schur complement we get
\begin{gather*}
    \begin{aligned}
        &\det(\mathbf{N}_{p+1}) = \det(A) \det \left( \mathbf{N}_{p} - \mathbbm{1} \mathbbm{1}^\intercal \otimes B A^{-1} B \right) \\
        &= \det(A) \det(A-B)^{p-1} \\
        &\qquad\cdot \det\big( A - B A^{-1} B + (p-1) (B - B A^{-1} B) \big),\\
        &= \det(A) \det(A-B)^{p-1} \\
        &\qquad\cdot \det\big( A-B + p BA^{-1}(A-B) \big),\\
         &= \det(A) \det(A-B)^{p}  \det\big( I + p BA^{-1} \big) ,\\
         &= \det(A-B)^{p} \det( A + p B ),\\
    \end{aligned}
\end{gather*}
where the second equality follows by applying the induction hypothesis to $\mathbf{N}_{p} - \mathbbm{1} \mathbbm{1}^\intercal \otimes B A^{-1} B$ and some algebraic manipulation. This completes the proof.\\

Part 2) From item \ref{item:tools_for_analysis.1} of the this lemma,
\begin{multline*}
    \det(\mathbf{N}_r - \lambda \mathrm{I}_r \otimes \mathrm{I})\\
    = \det(A- \lambda \mathrm{I} - B)^{r-1} \det(A -\lambda \mathrm{I} + (r-1) B),
\end{multline*}
implying that the spectrum of $\mathbf{N}_r$ coincides with that of $A-B$ and $A+(r-1)B$---modulo algebraic multiplicities.
Hence, $\mathbf{N}_r \succ 0$ results in $A - B \succ 0$ and $A + (r-1) B \succ 0$.
Furthermore, $\mathbf{N}_r \succ 0$ if and only if its principal submatrices are positive definite.
So, by applying the latter result to principal submatrices, we claim that $A + \ell B \succ 0$ for $\ell = 0,\cdots,r-2$.
Lastly, from item \ref{lem:algebraic_identities.1} of \Cref{lem:algebraic_identities}, for $\ell = 1,\cdots,r-1$ we have
\begin{gather*}
    A - \ell B \big( A + (\ell-1) B \big)^{-1} B \\
    =
    \big( A + \ell B \big) \big( A + (\ell-1) B \big)^{-1} \big( A - B \big),
\end{gather*}
which, by the first part of this claim, is invertible as a multiplication of invertible matrices.

 3) Since $\mathbf{N}_r$ and $A$ are invertible, the Schur complement $\mathbf{N}_{r-1} - L_{r-1} A^{-1} L_{r-1}^\intercal$ is also invertible where $L_{r-1} = \mathbbm{1}_{r-1}\otimes B$.
We prove the claim by induction on $r$.
For $r=2$, by \cite[Block matrix inverse forluma (0.8.5.6)]{horn2012matrix},
\begin{align*}
    \mathbf{N}_2^{-1} = \begin{bmatrix}
        H^{-1} & - H^{-1} B A^{-1} \\
        - A^{-1} B H^{-1} & A^{-1} + A^{-1} B H^{-1} B A^{-1}
    \end{bmatrix},
\end{align*}
where $H = A - B A^{-1} B$ is the Schur complement of $A$.
By \cite[Woodbury inversion formula (0.7.4.1)]{horn2012matrix}, $H^{-1} = A^{-1} + A^{-1} B H^{-1} B A^{-1}$, establishing the recurrence of diagonal blocks.
Also, $\mathbf{N}_2$ is symmetric, so is $\mathbf{N}_2^{-1}$ and thus establishing that $\mathbf{N}_2 \in \mathrm{PM}(2\times n, \mathbb{R})$.
Now, from \cref{lem:algebraic_identities.1} in \Cref{lem:algebraic_identities} with $n=1$, we get that
\begin{align*}
    A^{-1} B H^{-1} &= A^{-1} B \big( A - B \big)^{-1} A \big( A + B \big)^{-1} \\
    &= A^{-1} B \big( A + B \big)^{-1} A \big( A - B \big)^{-1} \\
    &= A^{-1} B \big( \mathrm{I} + A^{-1} B \big)^{-1} \big( A - B \big)^{-1} \\
    &= \big( \mathrm{I} - ( \mathrm{I} + A^{-1} B )^{-1} \big) \big( A - B \big)^{-1} = G_2,
\end{align*}
where we also used $\big( A - B \big)^{-1} A \big( A + B \big)^{-1} = \big( A + B \big)^{-1} A \big( A - B \big)^{-1}$ derived from \Cref{lem:algebraic_identities} \cref{lem:algebraic_identities.2}.
Hence,
\begin{gather*}
    \mathbf{N}_2^{-1} = \mathrm{I}_2 \otimes \big( H^{-1} + H^{-1} B A^{-1} \big) - \mathbbm{1}_2 \mathbbm{1}_2^\intercal \otimes \big( H^{-1} B A^{-1} \big).
\end{gather*}
Assume that the claim holds for $r=p$.
To extend the result to $r=p+1$, again  by \cite[Block matrix inverse forluma (0.8.5.6)]{horn2012matrix} and \cite[Woodbury inversion formula (0.7.4.1)]{horn2012matrix},
\begin{gather*}
    \mathbf{N}_{p+1}^{-1} = \begin{bmatrix}
        A & L_{p}^\intercal \\
        L_{p} & \mathbf{N}_p
    \end{bmatrix}^{-1}
    = \begin{bmatrix}
        P^{-1} & - P^{-1} L_{p}^\intercal \mathbf{N}_p^{-1} \\
        - \mathbf{N}_p^{-1} L_{p} P^{-1} & \big( \mathbf{N}_p - L_{p} A^{-1} L_{p} \big)^{-1}
    \end{bmatrix},
\end{gather*}
where $P = A - L_{p}^\intercal \mathbf{N}_p^{-1} L_{p}$ and $L_{p} = \mathbbm{1}_p \otimes B$.
Let $\mathbf{N}_p^{-1} = \mathrm{I}_p \otimes \big( F_p + G_p \big) - \mathbbm{1}_p \mathbbm{1}_p^\intercal \otimes G_p$ where,
\begin{align*}
    F_p &= \Big( A - (p-1) B \big( A + (p-2) B \big)^{-1} B \Big)^{-1}, \\
    G_p &= \big( A + (p-1) B \big)^{-1} B \big( A - B \big)^{-1},
\end{align*}
where the inversions are valid from item \ref{item:tools_for_analysis.2} of the current Lemma.
By simplification we get $P = A - p B \big( F_p - (p-1) G_p \big) B$ and from \Cref{lem:algebraic_identities} items \ref{lem:algebraic_identities.1} and \ref{lem:algebraic_identities.2},
\begin{gather*}
    \begin{aligned}
        F_p& - (p-1) G_p \\
        =& \Big( A - (p-1) B \big( A + (p-2) B \big)^{-1} B \Big)^{-1} \\
        &- (p-1) \big( A + (p-1) B \big)^{-1} B \big( A - B \big)^{-1} \\
        &= \big( A + (p-1) B \big)^{-1} \big( A + (p-2) B \big) \big( A - B \big)^{-1} \\
        &- (p-1) \big( A + (p-1) B \big)^{-1} B \big( A - B \big)^{-1} \\
        &= \big( A + (p-1) B \big)^{-1},
    \end{aligned}
\end{gather*}
where the first term in the second equation undergoes the first two items in \Cref{lem:algebraic_identities} consecutively.
The latter equality results in $P = A - p B \big( A + (p-1) B \big)^{-1} B$.
Next, considering the off-diagonal blocks of $\mathbf{N}_{p+1}^{-1}$, with some simplification, each block of $P^{-1} L_p^\intercal \mathbf{N}_p^{-1}$ is equivalent to $P^{-1} B \big( F_p - (p-1) G_p )$ and using the previous reasoning and \Cref{lem:algebraic_identities} it can be simplified to,
\begin{gather*}
    \begin{aligned}
        &P^{-1} B \big( F_p - (p-1) G_p )\\  
        &= P^{-1} B \big( A + (p-1) B \big)^{-1} \\ &= \big( A + p B \big)^{-1} \big( A + (p-1) B \big) \big( A - B \big)^{-1} B \big( A + (p-1) B \big)^{-1} \\
        &= \big( A + p B \big)^{-1} B \big( A - B \big)^{-1}.
    \end{aligned}
\end{gather*}
Similarly, each block of $\mathbf{N}_p^{-1} L_{p} P^{-1}$ is also equal to $\big( A + (p-1) B \big)^{-1} B \big( A - B \big)^{-1}$.
Therefore, it only remains to show that the blocks of $\big( \mathbf{N}_p - L_{p} A^{-1} L_{p} \big)^{-1}$ are consistent with the desired pattern in $\mathbf{N}_{p+1}^{-1}$.
Note that $\mathbf{N}_p - L_{p} A^{-1} L_{p} = \mathrm{I} \otimes \big( A - B \big) + \mathbbm{1} \mathbbm{1}^\intercal \otimes \big( B - B A^{-1} B \big)$.
Hence, with some algebraic rigor we can show that each diagonal term of $(\mathbf{N}_p - L_{p} A^{-1} L_{p})^{-1}$ is $\big( A + p B \big)^{-1} \big( A + (p-1) B \big) \big( A - B \big)^{-1}$
and each off-diagonal becomes $-\big( A + p B \big) B \big( A - B \big)^{-1}$.
Hence,
\begin{align*}
    \mathbf{N}_{p+1}^{-1} = \mathrm{I} \otimes \big( F_{p+1} + G_{p+1} \big) - \mathbbm{1} \mathbbm{1}^\intercal G_{p+1},
\end{align*}
with $F_{p+1}$ and $G_{p+1}$ defined as,
\begin{align*}
    F_{p+1} &= \big( A + p B \big)^{-1} \big( A + (p-1) B \big) \big( A - B \big)^{-1} \\
    G_{p+1} &= \big( A + p B \big) B \big( A - B \big)^{-1}.
\end{align*}

4)With direct multiplication and using the mixed-product property of Kronecker products,
\begin{gather*}
\begin{aligned}
    \mathbf{N}_r \mathbf{M}_r 
    =& \mathrm{I}_r \otimes \big( A - B \big) \big( C - D \big) + \mathbbm{1}_r \mathbbm{1}_r^\intercal \otimes B \big( C - D \big) \\
    &+ \mathbbm{1}_r \mathbbm{1}_r^\intercal \otimes \big( A - B \big) D + r \mathbbm{1}_r \mathbbm{1}_r^\intercal \otimes B D \\
    =& \mathrm{I}_r \otimes \big( A - B \big) \big( C - D \big) \\
    &+ \mathbbm{1}_r \mathbbm{1}_r^\intercal \otimes \Big( B \big( C - D \big) + \big( A - B \big) D + r B D \Big). \square
\end{aligned}
\end{gather*}





\bibliographystyle{ieeetr}
\bibliography{citations}

\begin{IEEEbiography}[{\includegraphics[width=1in,height=1.25in,clip,keepaspectratio]{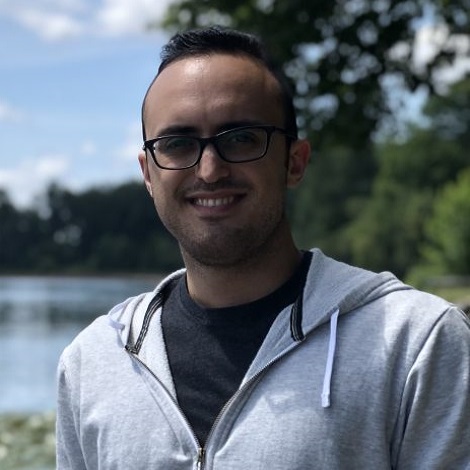}}]{Siavash Alemzadeh}
    (S'17) received his B.S. in Mechanical Engineering from Sharif University of Technology, Tehran, Iran in 2014. He received his M.S. in Mechanical Engineering in 2015 and his Ph.D. from William E. Boeing department of Aeronautics and Astronautics in 2020 from the University of Washington, Seattle, USA. He is currently a data and applied scientist at Microsoft.
    
    His research interests are reinforcement learning, data-driven control of distributed systems, and networked dynamical systems and their applications in transportation, infrastructure networks, and robotics.
\end{IEEEbiography}
\begin{IEEEbiography}[{\includegraphics[width=1in,height=1.3in,clip,keepaspectratio]{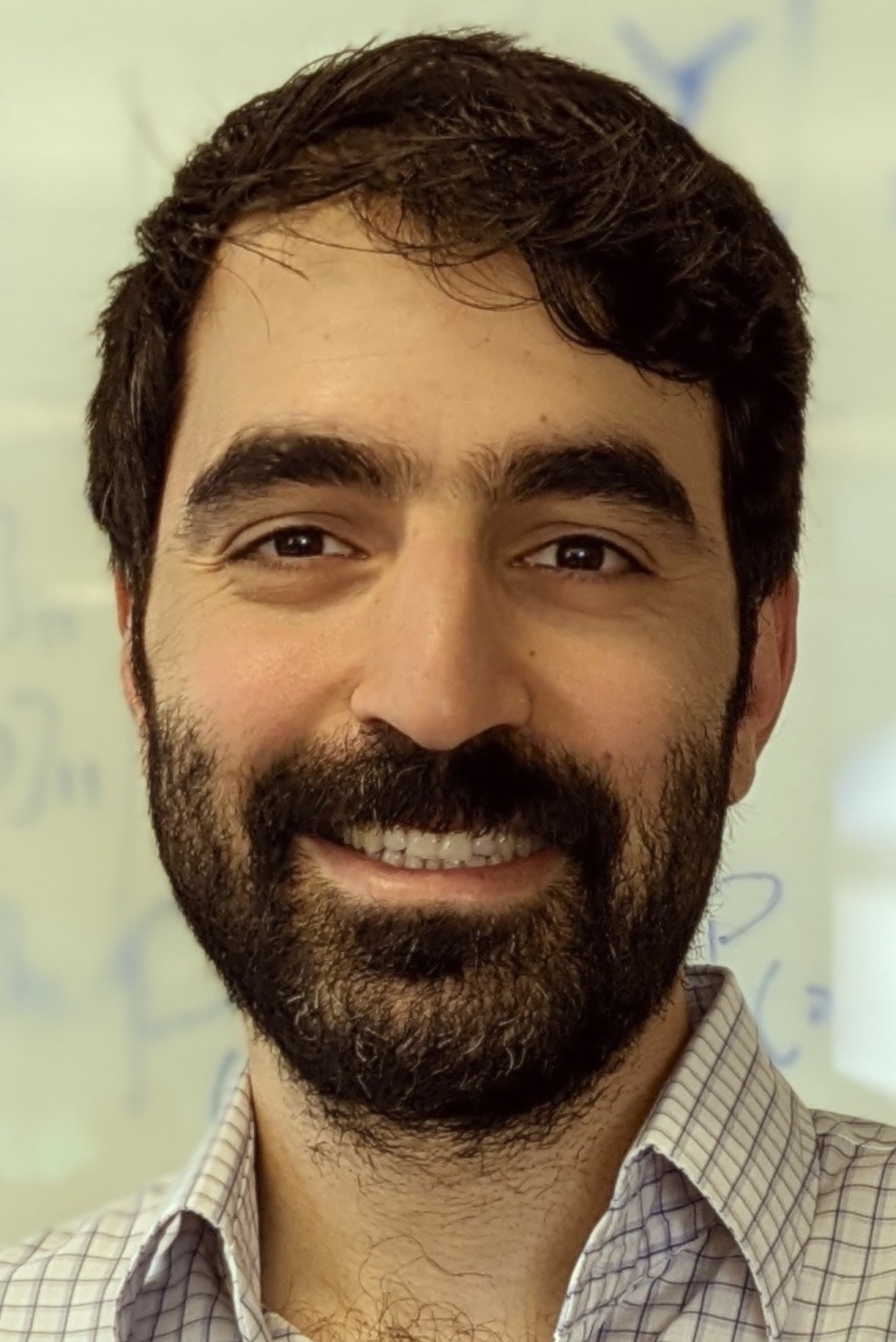}}]
{Shahriar Talebi} (Student Member, IEEE) 
received the Ph.D. degree in aeronautics and astronautics, specializing in control theory, and the M.Sc. degree in Mathematics, focusing on differential geometry, both from the University of Washington, Seattle, WA, USA, in 2023. He also received the B.Sc. degree from Sharif University of Technology, Tehran, Iran, in 2014, and the M.Sc. degree from the University of Central Florida, Orlando, FL, in 2017, both in electrical engineering. He is currently a postdoctoral fellow at the Harvard University with the School of Engineering and Applied Sciences (SEAS) and the Department of Statistics.
Dr. Talebi was the recipient of the 2022 Excellence in Teaching Award at UW. He is also a recipient of a number of scholarships, including the William E. Boeing Endowed Fellowship, Paul A. Carlstedt Endowment, and Latvian Arctic Pilot-A. Vagners Memorial Scholarship (UW, 2018–2019), as well as the Frank Hubbard Engineering Scholarship (UCF, 2017).

His research interests include control theory, differential geometry, learning for control, networked dynamical systems, and game theory.

\end{IEEEbiography}
\begin{IEEEbiography}[{\includegraphics[width=1in,height=1.25in,clip,keepaspectratio]{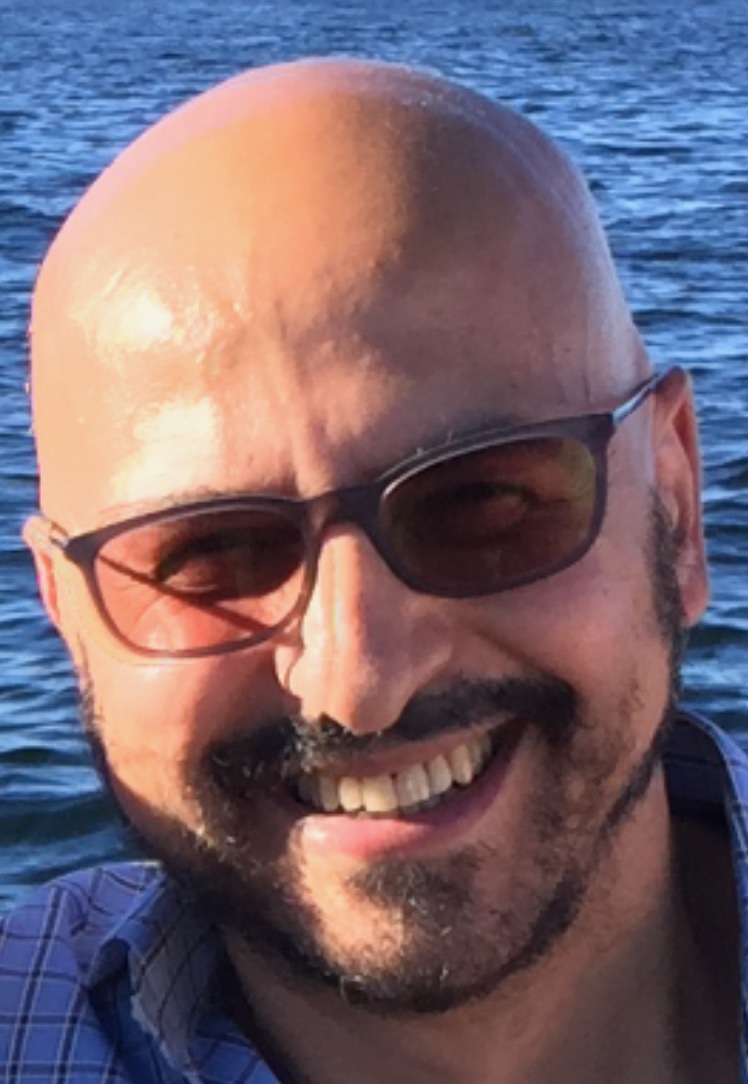}}]{Mehran Mesbahi}
    (F'15) received his Ph.D. from the University of Southern California, Los Angeles, in 1996. He was a member of the Guidance, Navigation, and Analysis group at Jet Propulsion Laboratory from 1996-2000 and an Assistant Professor of Aerospace Engineering and Mechanics at the University of Minnesota from 2000-2002.
    He is currently a Professor of Aeronautics and Astronautics, Adjunct Professor of Electrical and Computer Engineering and Mathematics at the
    University of Washington, Executive Director of the Joint Center for Aerospace Technology Innovation, and member of the Washington State Academy of Sciences.
    He was the recipient of NSF CAREER Award in 2001, NASA Space Act Award in 2004, UW Distinguished Teaching Award in 2005, and UW College of Engineering Innovator Award for Teaching in 2008.
    
    His research interests are distributed and networked aerospace systems, systems and control theory, and learning.
\end{IEEEbiography}

\end{document}